\documentclass[pdflatex,sn-mathphys-num]{sn-jnl}

\usepackage{graphicx}%
\usepackage{multirow}%
\usepackage{amsmath,amssymb,amsfonts}%
\usepackage{amsthm}%
\usepackage{mathrsfs}%
\usepackage[title]{appendix}%
\usepackage{xcolor}%
\usepackage{textcomp}%
\usepackage{manyfoot}%
\usepackage{booktabs}%
\usepackage{algorithm}%
\usepackage{algorithmicx}%
\usepackage{algpseudocode}%
\usepackage{listings}%
\usepackage{caption}
\usepackage{subcaption}

\newcommand{\fref}[1]{Fig. \ref{#1}}
\newcommand{\sref}[1]{Section \ref{#1}}

\newcommand{\aref}[1]{Appendix \ref{#1}}
\newcommand{\eref}[1]{eq. \ref{#1}}
\newcommand{\labelphantom}[1]{%
  \parbox{0pt}{\phantomsubcaption\label{#1}}%
}
\newcommand{\SigmaU}[0]{\mathbf{\Sigma}_{\mathcal{U}}}

\graphicspath{{figures/}}

\begin{document}

\title[]{Vibrational Entropy and Free Energy of Solid Lithium using Covariance of Atomic Displacements Enabled by Machine Learning}


\author*[1]{\fnm{Mgcini Keith} \sur{Phuthi}}\email{mkphuthi@umich.edu}

\author[2,3,4]{\fnm{Yang} \sur{Huang}}\email{hyhy123@ustc.edu.cn}

\author[2]{\fnm{Michael} \sur{Widom}}\email{widom@cmu.edu}

\author*[1]{\fnm{Venkatasubramanian} \sur{Viswanathan}}\email{venkvis@umich.edu}

\affil*[1]{\orgdiv{Mechanical Engineering}, \orgname{University of Michigan}, \orgaddress{\street{500 State St.}, \city{Ann Arbor}, \postcode{48105}, \state{Michigan}, \country{USA}}}

\affil[2]{\orgdiv{Department of Physics}, \orgname{Carnegie Mellon University}, \orgaddress{\street{5000 Forbes Ave.}, \city{Pittsburgh}, \postcode{15213}, \state{Pennsylvania}, \country{USA}}}

\affil[3]{
\orgname{University of Science and Technology of China},
\city{Hefei},
\postcode{230026},
\country{China}
}
\affil[4]{
\orgdiv{Suzhou Institute for Advanced Research},
\orgname{University of Science and Technology of China},
\city{Suzhou},
\postcode{215213},
\country{China}
}




\abstract{Vibrational properties of solids are key to determining stability, response and functionality. However, they are challenging to computationally predict at Ab-Initio accuracy, even for elemental systems. Ab-Initio methods for modeling atomic interactions are limited in the system sizes and simulation times that can be achieved. Due to these limitations, Machine Learning Interatomic Potentials (MLIPs) are gaining popularity and success as a faster, more scalable approach for modeling atomic interactions, potentially at Ab-Initio accuracy. Even with faster potentials, methodologies for predicting entropy, free energy and vibrational properties vary in accuracy, cost and difficulty to implement. Using the Covariance of Atomic Displacements (CAD) to predict entropy, free energy and finite-temperature phonon dispersions is a promising approach but thorough benchmarking has been hampered by the cost of Ab-Initio methods for sampling. In this work, we use a MLIP and the CAD to characterize the convergence of the predicted properties and determine optimal sampling strategies. We focus on solid lithium at zero pressure, showing that the MLIP-CAD approach reproduces experimental entropy, phonon dispersions and the martensitic transition while also comparing to more established methods.}

\keywords{Vibrational Entropy, Free Energy, Machine Learning Interatomic Potential, Lithium}



\maketitle


\section{Introduction}\label{sec1}
Computational materials discovery and characterization are powerful tools for engineering materials for critical applications such as sustainable energy \cite{jain_computational_2016, chen_critical_2020}. One of the most important properties of a material is the free energy landscape as a function of chemical composition, temperature and pressure. The free energy landscape determines the (meta-)stability of a material and is a strong indicator for synthesizability \cite{tszczypinski_can_2021}. After establishing stability, the material can be characterized for various other functional properties. There exist a number of methods of determining vibrational free energies depending on the material class. For solids, some of the most commonly used are Thermodynamic Integration (TI) \cite{frenkel_understanding_2001}, Temperature-Dependent Effective Potentials (TDEP) \cite{hellman_lattice_2011,hellman_temperature_2013} and the Quasiharmonic Approximation (QHA) \cite{korotaev_reproducibility_2018}. These methods employ varying degrees of approximation and can sometimes be difficult to implement for some material classes such as alloys \cite{grabowski_ab_2019}. Their feasibility is also determined by the implementation details or computational cost of the interatomic potential used to model the atomic interactions \cite{behler_machine_2021}.

A host of methods lie on the Pareto frontier of computational cost vs accuracy for modeling interatomic interactions, with some shown in \fref{fig:workflow}. Quantum chemistry methods are generally the most accurate but infeasible for most materials of interest \cite{martin_electronic_2008}. Density Functional Theory (DFT) is more feasible and extensively used due to explicitly including important quantum mechanical effects and for its generalizability across the entire periodic table  with acceptable accuracy for most applications\cite{martin_electronic_2008}. DFT however suffers from finite-size effects in simulations where long-range correlations matter since it scales poorly with the number of electrons ($\mathcal{O}(N^3)$) and is usually too slow for simulation protocols that require long timescale dynamics for the statistics of the calculated property to converge \cite{kresse_efficiency_1996}.

In cases where statistics and/or long-range correlations are significant, empirical potentials and Machine Learning Interatomic Potentials (MLIPs) fitted to experimental and/or ab-initio data are more suitable at the cost of some accuracy \cite{frenkel_understanding_2001}. The better scaling with the number of atoms ($\mathcal{O}(N)$) and orders of magnitude speed-up in property evaluation allows for collection of better statistics. Empirical potentials are however sometimes only qualitative in their predictions as they do not explicitly incorporate complicated quantum mechanical interactions. MLIPs on the other hand can be trained on ab-initio data to reproduce ab-initio predictions to remarkable accuracy \cite{kocer_neural_2022,behler_machine_2021}. They however suffer from poor generalizability in extrapolating outside the domain of the training data \cite{liu_discrepancies_2023, vita_data_2023}. MLIPs are therefore most reliable when predicting in-distribution properties in cases where collecting statistics (in size or simulation time) is the limiting factor for DFT or higher-order methods. The methodology described in this work takes advantage of the this ability to perform accurate in-distribution sampling.

Once an interatomic potential is chosen, a method needs to be selected to perform the vibrational free energy calculations. Thermodynamic Integration methods are often considered the gold standard as they are formally exact in predicting free energy differences \cite{korotaev_reproducibility_2018}. However, TI is generally challenging due to convergence issues and the need to identify an integration path for each system of interest. 
Some forms of TI require one to sequentially go from one ($T,P$) condition starting from the reference state making it impossible to fully parallelize across ($T,P$) space and thus computationally inefficient \cite{sluss_exploration_2022}. Hamiltonian Thermodynamic Integration can be parallelized across $T,P$ but requires long simulations where at each step contributions to the full Hamiltonian are turned on. Wang et al. performed HTI to compare different phases of lithium using a Deep Potential MLIP \cite{wang_data-driven_2023} where at each step in the integration path, 300,000 timesteps at 1fs intervals were used. 16 integration steps were used leading to 4,800,000 timesteps at each condition of temperature and pressure for each phase.

Alternatively, the Harmonic and Quasi-Harmonic Approximations are commonly used for solids. This is because the QHA is the simplest non-trivial approach to including effects of interacting phonons \cite{fultz_vibrational_2010,allen_theory_2020}. Coupled with the finite-displacement approach for predicting phonon spectra, the QHA is independent of the implementation of interatomic potential making it much easier to implement with MLIPs. Due to relatively low computational cost and feasibility of Density Functional Perturbation Theory, it is also used in Ab-Initio studies but is still costly for most cases \cite{korotaev_reproducibility_2018, jerabek_solving_2022}. When an MLIP is used, it only needs to be accurate in predicting the phonon spectrum of relevant phases at different pressures. The QHA explicitly ignores anharmonic contributions, except that of thermal expansion and its success hinges on the validity of the harmonic approximation in the material of interest or cancellation of errors. For highly anharmonic phases such as $\beta$-Ti or Zr, the QHA fails because the 0K phonon spectrum from which finite-temperature vibrational properties are derived has imaginary frequencies  \cite{huang_vibrational_2022,kadkhodaei_free_2017,hellman_lattice_2011}.

The Temperature Dependent Effective Potential (TDEP) and Covariance of Atomic Displacements (CAD) methods are effective phonon theories at finite temperatures \cite{hellman_lattice_2011,huang_vibrational_2022}. They both use statistics collected from MD simulations to construct ``effective'' force constant matrices giving phonon dispersion relations. The dispersion relations obtained are then used in standard harmonic formulae for entropy, free energy, heat capacity etc \cite{hellman_temperature_2013,huang_vibrational_2022}. The difference between the two methods is in how the force constant matrix is obtained. In TDEP, corrections to the force constant matrix of increasing order are obtained via an optimization procedure that minimizes the difference between a model Hamiltonian and the MD lattice dynamics. The TDEP method has been used in studies using AIMD, machine learning and empirical force fields  \cite{hellman_temperature_2013, romero_thermal_2015, grabowski_ab_2019}. In the CAD method, the force constant matrix is obtained by relating it to a matrix constructed using the covariance of atomic displacements from their ideal lattice sites. The derivation is briefly described in \sref{sec:cad_methods} and in more detail by Huang and Widom \cite{huang_vibrational_2022}. 

There also exist other approaches based on the Kirkwood approximation for entropy from MD simulations using pair correlation functions \cite{sluss_exploration_2022}. However some of these methods work best with pair-potentials, making it difficult to apply MLIPs.

In this work, we benchmark and discuss the combination of MLIPs with the CAD to provide computationally feasible and accurate predictions of vibrational properties of solids.  In particular, we investigate the entropy, phonon dispersion and free energy of lithium metal at finite temperatures for which a suitably accurate MLIP is available. By performing convergence tests and benchmarks, we discuss the choice of simulation protocols, parameters and potential pitfalls. These tests are much easier to perform with the cheaper interatomic potential. We show that the CAD method requires relatively small systems and significantly shorter simulation times than TI. Looking at the particular case of solid lithium, we compare CAD predictions with the QHA, TDEP and experimental results of thermodynamic properties and the lithium martensitic transition.



\section{Results}\label{sec2}
\begin{figure}[!tbh]
\centering
\includegraphics[width=\linewidth]{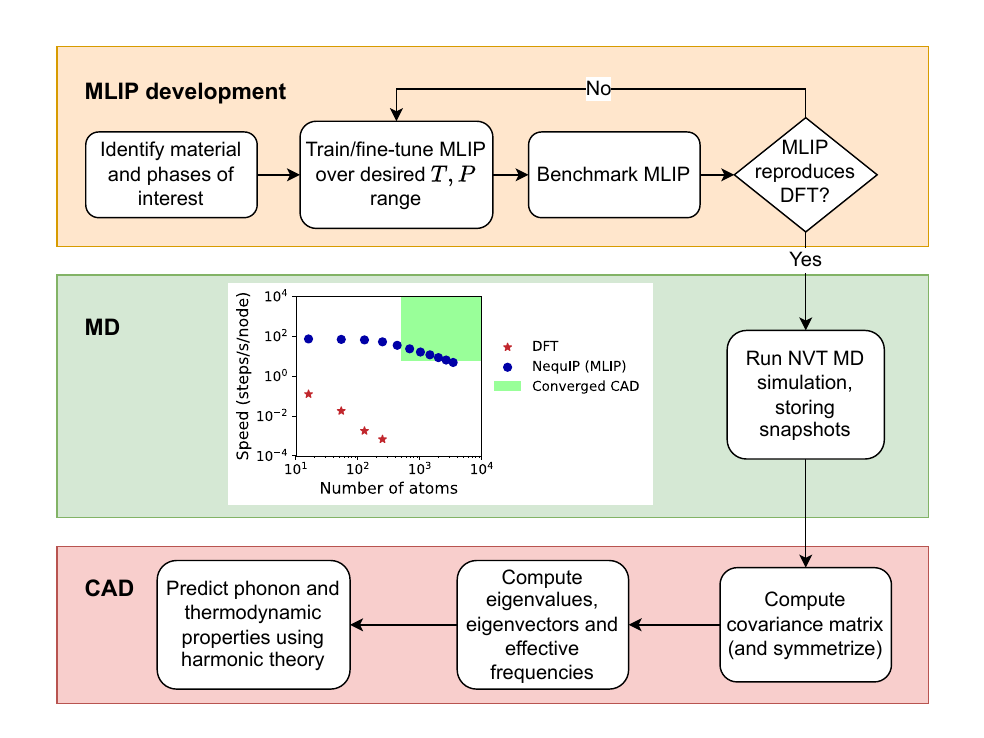}
\caption{Schematic workflow for CAD calculations with MLIPs. The inset plot demonstrates that to run a converged CAD calculation in less than an hour, it is necessary to use an MLIP such as NequIP or faster potential.}\label{fig:workflow}
\end{figure}

A workflow for computing vibrational properties of solids using CAD is given in \fref{fig:workflow}. The first step is to find or develop an appropriate atomistic potential for the system. In the case of MLIPs, a number of different methods exist such as generating datasets from scratch using active learning \cite{phuthi_accurate_2024,vandermause_--fly_2020} to fine-tuning pre-trained universal potentials \cite{deng_chgnet_2023,batatia_mace_2022}. The choice of MLIP depends on the accuracy and speed of performing MD. In \fref{fig:workflow}
, we also show that only MLIPs and and empirical potentials fall into the desirable window where we can perform converged CAD calculations within an hour per NVT simulation on a single high-performance computing node. Among the MLIPs and empirical potentials considered, the NequIP32 model is the most accurate. It is important to note that the Root Mean Square Error (RMSE) is not a complete indicator of accuracy. One should perform a number of different benchmarks on multiple properties to make sure the MLIP reproduces the DFT results it was trained on. The physics of the architecture is also important. The NequIP message passing neural netwrokd architecture implicitly incorporates very long-range interactions up to dozens of Angstroms generally resulting in better force predictions which govern the dynamics \cite{batzner_e3-equivariant_2022}.

Once the potential is determined, NVT simulations can be performed followed by CAD calculations as post-processing of MD snapshots. The MD and CAD portions of the workflow can be performed at any desired conditions in parallel.

By studying the phases of solid lithium, we quantify the convergence and uncertainties in the CAD calculated properties. Lithium is well-described by the QHA, having a Grunëisen parameter of 0.98 \cite{chuan-hui_gruneisen_2001} and it has a rich phase diagram with a temperature-driven martensitic transition from FCC to BCC at lower pressures \cite{jerabek_solving_2022}. We therefore compare the calculated properties to available experimental measurements and QHA predictions of thermodynamic properties.

\subsection{Convergence with system size and sampling time}

\begin{figure}[!tbh]
\centering
\labelphantom{fig:conv_size}
\labelphantom{fig:conv_steps}
\includegraphics[width=\linewidth]{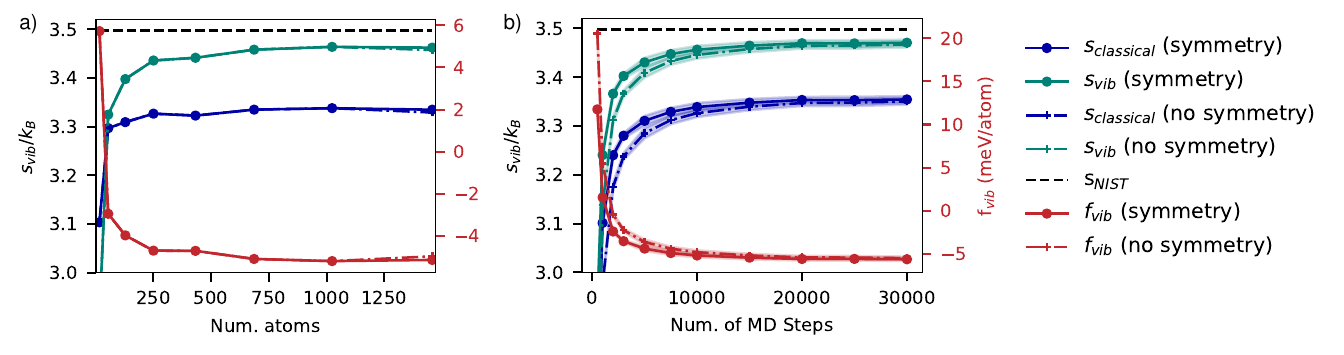}
\caption{Convergence of CAD entropy and vibrational free energy with and without symmeterization for BCC lithium at 300K with respect to a) number of atoms in unit cell using 10,000 timesteps b) number of MD steps using 432 atoms. Standard errors (too small to be visible) over five different sets of initial velocities are included as shaded regions. The black dashed line is the experimental value. Electronic entropy is not included for clarity and is very small at ${\sim}$0.04$k_B$.}\label{fig:convergence}
\end{figure}

Due to the stochastic nature of molecular dynamics simulations, it is important to quantify the convergence and statistical uncertainties of calculated properties with system size and simulation time. We determine this convergence by predicting entropy and vibrational free energy of BCC lithium at 300K. In \fref{fig:conv_steps}, we demonstrate the convergence criterion of the predicted entropy and free energy to be ${\sim}$20,000 timesteps, equivalent to 40ps sampled with a timestep of 2fs. To quantify uncertainty, we repeat each calculation 4 times with different random seeds for the initial velocities and include the standard deviation of results. The converged simulation time is significantly shorter than the simulation times needed in methods such as TI. We also discuss the effect of different sampling strategies such as using a sampling stride to reduce correlations, sampling later in the trajectory and varying the timestep in the Supplementary Information. All the strategies converge similarly to the same values as long as the simulation remains stable except in the case of phonon dispersion discussed in \sref{sec:phonons}. We also note that, for integrated properties such as entropy and free energy, using crystal symmetry to constrain CAD frequencies has little effect for sufficiently long simulations but drastically affects the phonon dispersion. The predictions using symmetry tend to converge slightly faster.

In \fref{fig:conv_size}, we observe that entropy and free energy values converge at ${\sim}$1,000 atoms. We further note that the free energy contributions $Ts$ and $f_{vib}$ at ${\sim}$500 atoms are 0.19meV/atom and 0.04meV/atom from the converged value respectively. These are small energy differences, well within typical error bounds of even typical DFT calculations. Therefore, in cases where high-throughput predictions are performed, it is acceptable to use ${\sim}$500 atoms to save computational effort. This also makes CAD more accessible to Ab-Initio Molecular Dynamics albeit still at significant cost.

For stable simulations, the predicted entropy and free energy are also robust as can be seen from the small standard deviation. The standard deviation with the random seed used to generate the initial velocities is small and barely visible in \fref{fig:convergence}. One therefore does not need to perform a large number of simulations with different initial velocities. Performing more than one simulation is however still useful as simulations in which instabilities or phase change occurs can be identified with large variations in entropy values. An example of diverging entropy and free energy values is shown in \fref{fig:deltag}, near the experimental melting point $T_m$=453.69K.

The determined convergence criteria mean that the vibrational entropy and free energy are easily accessible to MLIPs and empirical potential simulations. Each MD simulation at a given temperature and volume can be completed within 40min on a single NVIDIA A100 GPU using NequIP, one of the most costly MLIPs and orders of magnitude less for others. We also observe that the convergence criteria for system size and simulation times correspond similarly to those observed in the TDEP method \cite{hellman_temperature_2013}.

\subsection{Finite-Temperature Phonon Dispersion Relations}\label{sec:phonons}

\begin{figure}[!tbh]
\centering
\includegraphics[width=0.8\linewidth]{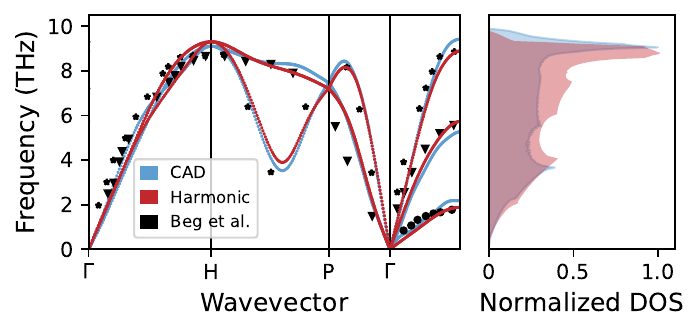}
\caption{Phonon dispersions and DOS for BCC lithium at 300K predicted using the CAD and the harmonic approximation at the corresponding equilibrium volume determined by NPT simulations. Both approaches reproduce experimental results well.}\label{fig:phonons}
\end{figure}

The formulation of the CAD method in terms of effective harmonic frequencies naturally allows the prediction of phonon dispersion relations and density of states (DOS). In \fref{fig:phonons}, we compare an example of a CAD dispersion and DOS for BCC lithium at 300K to experimental dispersion measured at the same temperature. Apparently, the CAD dispersion reproduces the experimental results very well. It particularly captures the softening of phonon modes as temperature increases, for example at the $H$ point resulting in a decreasing maximum frequency with temperature compared corresponding to a smaller lattice parameter. A traditional harmonic phonon calculation at the predicted equilibrium volume corresponding to $T$=300K is an alternative approach to obtaining a ``finite-temperature'' dispersion. The harmonic dispersion also corresponds very well with experiments. This highlights the consistency between the two different approaches to obtaining the dispersion relations albeit the harmonic calculation being significantly cheaper since it does not require MD simulations. 

When predicting the CAD dispersion and DOS, there are a few notes of caution. The CAD dispersion converges after slightly longer simulation time than entropy, but not necessarily needing more timesteps. It is usually better to run a simulation with larger timesteps, even as much as 5fs. Additionally, symmeterization of the covariance matrix is critically important for predicting a reasonable CAD dispersion even if the entropy and free energy are not as affected. Instabilities in the simulation can lead to ``oversoftening'' of some phonon modes and even imaginary frequencies. The physical significance of these imaginary frequencies should be considered with care as the frequencies in CAD are empirical. Multiple repeated runs with different initial velocities and larger supercells give a clearer picture of the degree of variation in the CAD dispersion and DOS.

\subsection{Thermodynamics at Zero Pressure}

\begin{figure}[!htb]
\centering
\labelphantom{fig:thermal_expansion}
\labelphantom{fig:cp}
\labelphantom{fig:s}
\labelphantom{fig:s_error}
\labelphantom{fig:g}
\labelphantom{fig:deltag}
\includegraphics[width=0.9\linewidth]{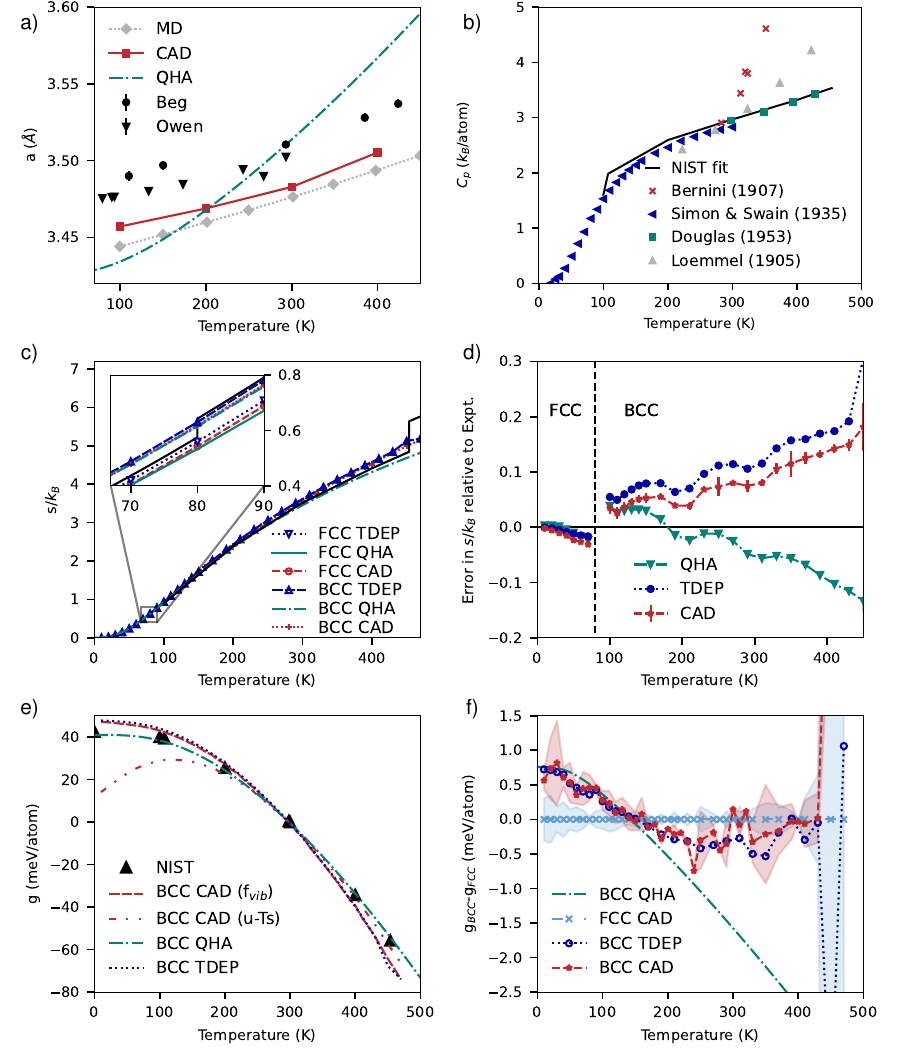}
\caption{Thermodynamic properties as a function of temperature, namely a) Lattice constant b) Electronic entropy c) Total Entropy of FCC and BCC, including electronic entropy, d) Residual of entropy with respect to NIST experimental data. All methods are in good agreement with experimental data and each other below 100K, capturing the small discontinuity $\Delta s$=0.067$k_B$ within 5\%. Above 100K, CAD, TDEP and QHA do well and are well within experimental error but diverge in different directions. e) Gibbs free energy referenced at $T=$298.15K. CAD and TDEP predictions are slightly different from QHA and experiment which agree well. f) Difference between the FCC and BCC Gibbs free energy. All methods predict $T_c{\sim}$150K compared to 80K from experiment. The shaded area is the standard error for each phase in CAD. }\label{fig:zerogpa}
\end{figure}

\nocite{beg_temperature_1976}
\nocite{owen_x-ray_1954}

We show some thermodynamic properties of lithium metal as a function of temperature at zero pressure, effectively corresponding to room pressure in \fref{fig:zerogpa}. For CAD, each calculation is performed at the corresponding equilibrium volume at the given temperature found by averaging the volume in an NPT simulation. Therefore thermal expansion due to classical dynamics is accounted for. 

In \fref{fig:thermal_expansion}, we show that the lattice constant as predicted by the volume that minimizes the CAD free energy is most consistent with experimental data. The equilibrium volume as predicted by NPT simulations in MD also have matching qualitative behavior. The CAD lattice constant is closer to experimental values and always higher than the MD lattice constant as it accounts for the zero-point volume correction. The difference between them however is small when considering it's effect on predicted entropies and free energies. The QHA predicted lattice constant deviates significantly more from the trend in experimental data despite matching the experimental room temperature lattice constant very well.

In \fref{fig:s} we show the entropy as predicted by CAD, TDEP and QHA compared to experiment including the small linear correction of electronic entropy shown in the Supplementary Information. At low temperatures, it is expected that the QHA performs well as it is a low temperature approximation. CAD and TDEP perform just as well at low temperatures as the dynamics converge to becoming more harmonic in the low-temperature regime. Figure \ref{fig:s} also shows the entropy of BCC lithium, including electronic entropy up to the melting point, $T_m$=453.69K. Both CAD and QHA describe the system within experimental error bounds (${\sim} 0.05k_B$) of the NIST data up to the melting point \cite{chase_nist-janaf_1998}. We also highlight that there is significant variation in experimental heat capacity ($C_p$) measurements near the melting point as shown in \fref{fig:cp}. The NIST data is fit to the most recent Douglas et al. \cite{douglas_lithium_1955} experiments. The Douglas et al. results are the lowest values measured, all other experiments measured higher heat capacity by up to 30\% which would result in higher entropy above 300K. More experiments with modern techniques would be useful to investigate the discrepancy between different experiments. CAD and TDEP generally estimate higher entropy values at higher temperatures than QHA which could explain the higher experimental values of heat capacity and can potentially be attributed to anharmonicity beyond the QHA. In all properties shown, there is a clear correspondence between the TDEP and CAD predicted quantities since they both employ classical (Born-Oppenheimer) molecular dynamics to sample the configuration space.

In \fref{fig:g}, we show the Gibbs free energy of BCC lithium as a function of temperature. The CAD and TDEP predictions decrease faster with temperature whereas the QHA prediction matches very well with experiment. This difference is attributed to the slightly higher entropy in \fref{fig:s}. In this case, it appears the QHA performs significantly better. The CAD and TDEP free energy shown is based on \eref{eq:helmholtz} which is a harmonic equation but using CAD frequencies. Korotaev et al. note that using the harmonic expression for the free energy is questionable even in the TDEP approach and argues that it is better to use 

\begin{equation}\label{eq:uts}
    f = u - Ts,
\end{equation}

where $u$ is the average internal energy from the molecular dynamics simulation and $s$ is the TDEP entropy. In the case of lithium at low temperature however, such an approximation leads to a gross violation of the second law of thermodynamics as shown by the non-monotonic free energy as a function of temperature in \fref{fig:g}. One potential cause of this issue is in the inability of the MLIP to perfectly reproduce DFT. Furthermore, the underlying DFT used to train the MLIP also has an intrinsic error from parameter choices, choice of exchange-correlation functional and other approximations \cite{sholl_density_2009}. These differences however may not account for why the QHA performs so well for the free energy. 

We believe the most likely source source of error is the assumption of Born-Oppenheimer dynamics in the low temperature limit for the very light lithium atoms. The Born-Oppenheimer approximation ignores nuclear quantum effects such as coherence \cite{markland_nuclear_2018}. The assumption of Boltzmann distributed velocities and classical dynamics is itself flawed at low temperatures which are better described with a full quantum mechanical treatment. For instance, one can calculate the thermal De Broglie wavelength for lithium at 10K as ${\sim} 2\AA{}$, significantly larger than the atomic displacements. The effect is to significantly overestimate the enthalpy at low temperatures which amounts by assuming that the equipartition of energy holds which is invalid at low temperatures \cite{frenkel_understanding_2001}. Feng et al. found that nuclear quantum effects reduced the free energy of the high pressure cI16 phase of lithium by up to 15meV/atom at nearly 300K using Path-Integral Molecular Dynamics \cite{markland_nuclear_2018}. It would therefore be useful to either improve or rigorously validate the use of harmonic expressions using CAD or TDEP frequencies to identify the source of error. More discussion on this issue is included in the Supplementary Information.

An alternative approach might be to employ Wigner's quantum correction to classical statistical mechanics \cite{wigner_quantum_1932}. This is a rapidly convergent perturbative expansion of the free energy in the ratio of $\hbar\omega/k_B T$ \cite{deserno_leading_2023}. 

We also note that using \eref{eq:uts} apparently gives better agreement with experiment at higher temperature explaining how it has been successfully used to model other transitions such as that of Ti which occurs at 1155.15K where a classical approach might be more appropriate \cite{huang_ab_2022}.

Free energy calculations are most useful if used to determine phase stability and phase transition critical temperatures ($T_c$). Experimentally, at $T_c$=80($\pm 10$)K, a martensitic transition from the face-centered cubic (FCC) structure to the body-centered cubic (BCC) structure occurs with increasing temperature \cite{hultgren_selected_1963}. 

In \fref{fig:deltag} we show the Gibbs free energy difference between FCC and BCC lithium as a function of temperature. We observe that CAD, TDEP and QHA methods predict a phase transition at about 150K where the free energies cross, significantly higher than the experimental value. Predicting $T_c$ for this transition accurately has been a long standing challenge for computational methods. The free energy differences associated with the transition are of the order of 1meV/atom. An error in the cohesive energy of 0.5meV/atom can change $T_c$ by up to 100K. The NequIP MLIP has a reported RMSE of 1meV/atom for energy. The effect of the zero-point correction to the volume, a purely quantum mechanical effect plays a small role. We estimate that this effect contributes at most a 0.1meV/atom difference. An additional source of error is the choice of Exchange-Correlation Functional. We found that while PBE and PBESol gave very similar results, SCAN gave a cohesive energy about 0.3meV/atom higher. 

In \fref{fig:deltag}, it appears as though the FCC and BCC phases have very similar Gibbs free energy above $T_c$ in the CAD prediction, unlike in the QHA before a large divergence. The large divergence is due to melting which can be observed in the atomic configurations and happens for FCC before it happens for BCC. The divergence and deviation of the CAD entropy may be correlated with the (meta-)stability of the phase being simulated, even if it remains the dominant phase in the simulation. This would mean that if one is not careful, it is possible to overestimate an unstable structure's entropy. An overestimated entropy leads to a more negative contribution of the $Ts$ term to the free energy. Ab-initio calculations by Jerabek et al.\cite{jerabek_solving_2022} and Xie et al. \cite{xie_origin_2008} argued that the FCC and BCC phase are quasi-degenerate with BCC being metastable below the $T_c$ with barriers less than 1meV/atom corresponding to ${\sim}$12K. The Zero Point Energy (ZPE) and vibrational entropy are important in stabilizing BCC as it is a temperature-driven transition \cite{schaeffer_boundaries_2015}. With increasing temperature, the transition from FCC to BCC becomes ``almost barrierless'' hence FCC is not metastable above $T_c$. This conclusion aligns with the observation in \fref{fig:deltag}.

These subtle caveats imply that confident use of the CAD method requires one to already know the stable phase. There are however some simple checks one can make to test the entropy prediction such as checking if the magnitude of the entropy is reasonable compared to other phases at the same temperature, if the dependence of free energy with temperature is strictly decreasing and more weakly, if the entropy decreases with increasing pressure if the material does not have negative thermal expansion \cite{liu_zentropy_2022}. These constraints are placed by the second law of thermodynamics and are discussed in \aref{app:dgdt}. This also means the CAD entropy is a potential descriptor for meta-stability since divergence of the entropy suggests there exists a possible phase transition that may not be apparent in structural analysis or requires too long a simulation time to occur.

In the entropy, free energy and phase transition predictions, the CAD and TDEP approaches have very similar results. This suggests that the limitations observed are likely limitations to either the grounding of the methods in the harmonic theory, or the use of classical molecular dynamics simulations for a strongly quantum system. And so far, the two methods appear interchangeable for the case of lithium metal.

\section{Discussion}\label{sec12}

The CAD method overcomes some of the computational challenges of TI and in combination with MLIPs, is significantly computationally cheaper. Firstly, CAD properties (vibrational entropy, free energy etc.) can be spot-evaluated at a particular temperature and pressure for a particular phase. This means the evaluations are completely parallelizable in $T,P$ space as with HTI and TDEP. However, one only needs ${\sim} 20,000$ samples from a single NVT simulation, ${{\sim}} 240$ times less timesteps at each $T,P$ evaluation compared to HTI. Perhaps most importantly, since CAD is implemented as a post-processing step to an NVT/NPT simulation it is significantly easier to use. The researcher does not need to spend significant effort determining parameters like integration steps, and reference states. Unlike TI however, the CAD method only accounts for vibrational properties. Liquid, configurational entropy and free energy would have to be treated within a different framework. 

Compared to CAD, the QHA is significantly computationally cheaper when using a MLIP. This is because in the QHA, only the calculation of the phonon dispersion requires the interatomic potential at each pressure. Subsequent steps are mostly post-processing using standard analytical formulae for the temperature dependence. The CAD method, like TDEP requires MD simulations at each temperature. The CAD method however does not fail for anharmonic materials at 0K unlike the QHA. Huang and Widom showed that the entropy and phase transition of $\beta$-Ti can be predicted using the CAD method. Additionally, CAD implicitly accounts for anharmonicity as it is an effective harmonic theory, with the expectation of better results at higher temperatures compared to QHA. As lithium does not have strong anharmonicity, studies on other materials would have to be carried out to validate this assertion.

\section{Methods}\label{sec11}

In this section, we describe the basics of the theory behind the CAD calculations, model architectures, data generation and simulation details. 

\subsection{Covariance of Atomic Displacements}\label{sec:cad_methods}
The CAD method introduced by Huang and Widom\cite{huang_vibrational_2022} allows the prediction of vibrational properties such as entropy and free energy of solids at finite temperatures, and phonon dispersion including anharmonicity. The vibrational properties are derived from a covariance matrix ($\SigmaU$) constructed from molecular dynamics or Monte-Carlo simulations. Each matrix element of $\SigmaU$ is given by the covariance between the displacement ($x_{i\mu}$) of atom $i$ in each coordinate $\mu$ and the displacement ($x_{j\nu}$) of every other atom $j$ in each coordinate $\nu$ relative to their ideal lattice positions over the course of a trajectory. $\SigmaU$ is therefore a $3N\times3N$ matrix with elements $\Sigma_{i\mu j\nu} = \langle x_{i\mu}x_{j\nu} \rangle $.
One can relate the mass-weighted covariance matrix, $\Tilde{\Sigma}_{i\mu,j\nu}=\Sigma_{i\mu,j\nu}\sqrt{m_im_j}$ to the mass-reduced force constant matrix $\Tilde{C}_{i\mu,j\nu}=C_{i\mu,j\nu}/\sqrt{m_im_j}$ using the relation

\begin{equation}\label{eq:force_const}
    \Tilde{C} = \frac{1}{\beta} \Tilde{\Sigma}^{-1}.
\end{equation}
Equation \ref{eq:force_const} identifies the eigenvalues of $\Tilde{C}$ as inverse frequencies $\omega_{k\mu}^{-1}$ which share eigenvectors with $\Tilde{\Sigma}$. For a harmonic potential \eref{eq:force_const} is exact whereas for a general potential, it is an \textit{effective} force constant matrix from which to extract effective vibrational frequencies that include anharmonic effects. The frequencies are effective in that they implicitly account for anharmonicity beyond the quasiharmonic approximation which only accounts for anharmonicity due to thermal expansion. The expectation is therefore that the CAD frequencies represent higher temperature properties more accurately.

Given the set of CAD frequencies, it is therefore possible to use the standard harmonic expressions for thermodynamic properties. The vibrational entropy at temperature $T$ is given by
\begin{equation}\label{eq:s_quantum}
    s_{vib} = k_B\sum\limits_{k\mu}\left[\frac{\beta\hbar\omega_{k\mu}}{e^{\beta\hbar\omega_{k\mu}}-1}-\ln(1-e^{-\beta\hbar\omega_{k\mu}})\right],
\end{equation}
where $k_B$ is the Boltzmann constant, $\beta=1/k_BT$ and $\hbar$ is the reduced Planck's constant, $e$ is the exponential factor and $\Lambda$ is the thermal de Broglie wavelength of the material.

An alternative expression for the entropy of a single atomic species can be written succinctly in terms of the covariance matrix as
\begin{equation}\label{eq:s_classical}
    s^{classical}_{vib} = \frac{1}{2}k_B\ln\left[
    (2\pi e/\Lambda^2)^{3N}\det{(\SigmaU)}\right]+\frac{3}{2}k_B,
\end{equation}
a form that does not explicitly invoke vibrational frequencies or the harmonic approximation. The expression in \eref{eq:s_classical} is termed the classical entropy as it is derived from information-theoretic principles but diverges to $-\infty$ as $T\rightarrow 0$. The third law of thermodynamics demands that the entropy should vanish instead. We therefore use \eref{eq:s_quantum} unless otherwise stated as it more faithfully reproduces experimental results over the entire temperature range as shown in \fref{fig:convergence} and does not erroneously diverge.

The vibrational Helmholtz free energy can be expressed as

\begin{equation}\label{eq:helmholtz}
    f_{vib} = \frac{1}{N}\sum\limits_{k\mu}\left[\frac{1}{2}\hbar\omega_{k\mu} + \frac{1}{\beta}\ln\left( 1-e^{-\beta\hbar\omega_{k\mu}}\right)\right].
\end{equation}

Other properties such as vibrational heat capacity, thermal conductivity, can also be derived using standard formulae. Additionally, given the frequencies and eigenvectors of the force constant matrix, one can also construct finite-temperature phonon dispersion curves and the density of states (DOS).

The total free energy per atom ($f(v,T)$) of a single element system depends on specific volume ($v$) and is given by

\begin{equation}\label{helmholtz}
    f(v,T) = u(v,T=0) + f_{vib}(v,T) + f_{el}(v,T).
\end{equation}

$u(v, T=0)$ is the potential energy per atom of the ideal crystal structure. We reference all potential energies relative to that of FCC lithium in equilibrium at 0K. $u$ is obtained from the chosen interatomic potential which in this case is the NequIP MLIP. It is also reasonable to use the DFT calculated $u$ for the static cell since it is cheap to calculate and more reliable. $f_{el}$ is the electronic free energy which can be approximated from DFT using Fermi-Dirac smearing at the corresponding temperature. 

We are usually interested in the stability of materials under conditions of pressure and temperature hence the Gibbs free energy ($g(p,T)$) or similarly the NPT ensemble is the more appropriate thermodynamic potential. $g$ can be calculated from $f$ in the canonical NVT ensemble or enthalpy ($h$) in the NPS ensemble via Legendre transforms given by

\begin{equation}\label{eq:g1}
    g(P,T) = f(v,T) + Pv = h(P,s) - Ts
\end{equation}
We work under the assumption that we can perform simulations in the more convenient canonical NVT ensemble and use eq. \ref{eq:g1} to get $g$. This assumption is exact in the thermodynamic limit and is already implicitly assumed in MD when we approximate the true thermodynamic quantities as the mean values from simulations \cite{frenkel_appendix_2002}. We demonstrate the consistency of using the NVT simulation in place of the NPT simulation in \fref{fig:thermal_expansion} which shows that the equilibrium volume obtained using an NPT simulation corresponds well with the volume that minimizes $f(v,T)$ obtained from multiple NVT simulations for a given $T$. Expressed in terms of values we extract from simulation, 

\begin{equation}\label{eq:g}
    g(P,T) = u(v) + f_{vib}(v,T) + f_{el}(T).
\end{equation}

The most stable structure out of a set of candidates at given temperature and pressure is the one that minimizes eq. \ref{eq:g}.

\subsection{NequIP Potential}
The MLIP used in this work was developed by Phuthi et al. and details of the dataset and architecture are available \cite{phuthi_accurate_2024}. This MLIP was shown to have high accuracy for describing a number of properties of lithium metal. The NequIP architecture which uses equivariant features of rotation order up to 2 is known to be very data efficient and performs with state-of-the-art accuracy on a number of benchmarks \cite{phuthi_accurate_2024,batzner_e3-equivariant_2022}.

\subsection{Simulation Parameters}
All DFT, MD and CAD simulations are managed within our open-source simulation framework, Atomic Simulation Tools (ASIMTools). The framework allows the building of simulation protocols incorporating multiple different software tools, allowing reproducible simulations. 

For DFT calculations, parameters were converged to less than 1meV/atom in energy. DFT calculations were done using Quantum Espresso \cite{giannozzi_quantum_2009} within the Generalized Gradient Approximation using the Perdew-Burke-Eizenhoff exchange correlation functional \cite{perdew_generalized_1996} and the Projector Augmented Wave approach \cite{blochl_projector_1994} with a plane wave cutoff energy of 1360eV. We used the pseudopotential Li.pbe-s-kjpaw\_psl.1.0.0.UPF from
http://www.quantum-espresso.org. A uniform Brillouin Zone spacing of 0.02$\AA{}^{-1}$ with a Monkhorst-Pack \cite{monkhorst_special_1976} sampling procedure was used. To help with convergence of the the Fermi surface, Methfessel-Paxton \cite{methfessel_high-precision_1989} smearing using a smearing width of 0.27eV was chosen.

For TDEP calculations, the same trajectories were used as in the CAD calculations and only up to second-order force constants were included.

\subsection{Molecular Dynamics details}
Molecular Dynamics simulations were performed in the Large-scale Atomic/Molecular Massively Parallel Simulator (LAMMPS) \cite{thompson_lammps_2022}. Unless otherwise stated, simulations were performed using a timestep of 2fs. Isobaric (NPT) or isothermal (NVT) simulations were performed using Nos\'e-Hoover style barostat and thermostat equations of motion by Shinoda et al. \cite{shinoda_rapid_2004}. For CAD runs, 20,000 timesteps were used unless stated otherwise. For FCC, supercells with 500 atoms and for BCC, supercells with 432 atoms were used unless stated otherwise.







\backmatter

\bmhead{Supplementary information}
The provided supplementary information provides additional benchmarks for the CAD method and sampling strategies as well as more details of the simulations performed.

\bmhead{Acknowledgements}
This work was supported in part by Oracle Cloud credits and related resources provided by Oracle for Research. MW and YH were supported by DOE grant DE-SC0014506.

\section*{Declarations}

\subsection{Data and Code Availability}
The MLIP and training data used in this work is available on Zenodo \cite{mgcini_dataset_2024}. The code for the CAD post-processing is available from https://alloy.phys.cmu.edu/software/CAD.html. An interface for running benchmarks, MD (using LAMMPS or ASE) and CAD altogether is available as an extension to ASIMTools and will be made available on github. For all TDEP calculations, the open-source TDEP software by Knoop et al. was used \cite{hellman_temperature_2013,knoop_tdep_2024}. 

\subsection{Conflicts of Interest}
The authors declare no conflicts of interest.

\subsection{Author Contribution}
Conceptualization, M.K.P; Methodology - Modeling - M.K.P., Y.H., M.W.; Investigation, M.K.P.; Formal Analysis, M.K.P., M.W.; Data Curation, M.K.P. ; Writing - Original Draft, M.K.P., M.W. and V.V.; Writing - Review \& Editing - M.K.P.; Supervision. M.W. and V.V.








\begin{appendices}

\section{}\label{app:dgdt}

We briefly discuss why the second law of thermodynamics imposes that the Gibbs free energy at constant pressure must be a decreasing function of temperature. One can write out the differential form of the Gibbs free energy

\begin{equation}
    dG = -SdT + Vdp + \mu dN
\end{equation}

where 
\begin{equation}\label{eq:dgdt}
    \frac{\partial G}{\partial T}\big|_{p,N} = -S < 0.
\end{equation}

Since the entropy $S$ is positive, $G$ can only be a decreasing function of $T$.

Now consider the dependence of entropy on pressure at constant temperature i.e.

\begin{equation}
    \frac{\partial S}{\partial p}\bigg|_{T,N} = -\frac{\partial V}{\partial T}\bigg|_{p,N}
\end{equation}

The right-hand side is derived using a Maxwell relation and is related to the coefficient of thermal expansion ($\alpha$) by

\begin{equation}
    \alpha = \frac{1}{V}\left(\frac{\partial V}{\partial T}\bigg|_p\right).
\end{equation}

For most materials, the thermal expansion coefficient is positive but there are a few exceptions such as pure silicon. One can therefore not rule out that a materials with increasing entropy as a function of pressure is unstable but it is likely.




\end{appendices}


\bibliography{sn-bibliography}


\begin{thebibliography}{51}
\ifx \bisbn   \undefined \def \bisbn  #1{ISBN #1}\fi
\ifx \binits  \undefined \def \binits#1{#1}\fi
\ifx \bauthor  \undefined \def \bauthor#1{#1}\fi
\ifx \batitle  \undefined \def \batitle#1{#1}\fi
\ifx \bjtitle  \undefined \def \bjtitle#1{#1}\fi
\ifx \bvolume  \undefined \def \bvolume#1{\textbf{#1}}\fi
\ifx \byear  \undefined \def \byear#1{#1}\fi
\ifx \bissue  \undefined \def \bissue#1{#1}\fi
\ifx \bfpage  \undefined \def \bfpage#1{#1}\fi
\ifx \blpage  \undefined \def \blpage #1{#1}\fi
\ifx \burl  \undefined \def \burl#1{\textsf{#1}}\fi
\ifx \doiurl  \undefined \def \doiurl#1{\url{https://doi.org/#1}}\fi
\ifx \betal  \undefined \def \betal{\textit{et al.}}\fi
\ifx \binstitute  \undefined \def \binstitute#1{#1}\fi
\ifx \binstitutionaled  \undefined \def \binstitutionaled#1{#1}\fi
\ifx \bctitle  \undefined \def \bctitle#1{#1}\fi
\ifx \beditor  \undefined \def \beditor#1{#1}\fi
\ifx \bpublisher  \undefined \def \bpublisher#1{#1}\fi
\ifx \bbtitle  \undefined \def \bbtitle#1{#1}\fi
\ifx \bedition  \undefined \def \bedition#1{#1}\fi
\ifx \bseriesno  \undefined \def \bseriesno#1{#1}\fi
\ifx \blocation  \undefined \def \blocation#1{#1}\fi
\ifx \bsertitle  \undefined \def \bsertitle#1{#1}\fi
\ifx \bsnm \undefined \def \bsnm#1{#1}\fi
\ifx \bsuffix \undefined \def \bsuffix#1{#1}\fi
\ifx \bparticle \undefined \def \bparticle#1{#1}\fi
\ifx \barticle \undefined \def \barticle#1{#1}\fi
\bibcommenthead
\ifx \bconfdate \undefined \def \bconfdate #1{#1}\fi
\ifx \botherref \undefined \def \botherref #1{#1}\fi
\ifx \url \undefined \def \url#1{\textsf{#1}}\fi
\ifx \bchapter \undefined \def \bchapter#1{#1}\fi
\ifx \bbook \undefined \def \bbook#1{#1}\fi
\ifx \bcomment \undefined \def \bcomment#1{#1}\fi
\ifx \oauthor \undefined \def \oauthor#1{#1}\fi
\ifx \citeauthoryear \undefined \def \citeauthoryear#1{#1}\fi
\ifx \endbibitem  \undefined \def \endbibitem {}\fi
\ifx \bconflocation  \undefined \def \bconflocation#1{#1}\fi
\ifx \arxivurl  \undefined \def \arxivurl#1{\textsf{#1}}\fi
\csname PreBibitemsHook\endcsname

\bibitem[\protect\citeauthoryear{Jain et~al.}{2016}]{jain_computational_2016}
\begin{barticle}
\bauthor{\bsnm{Jain}, \binits{A.}},
\bauthor{\bsnm{Shin}, \binits{Y.}},
\bauthor{\bsnm{Persson}, \binits{K.A.}}:
\batitle{Computational predictions of energy materials using density functional theory}.
\bjtitle{Nature Reviews Materials}
\bvolume{1}(\bissue{1}),
\bfpage{1}--\blpage{13}
(\byear{2016})
\doiurl{10.1038/natrevmats.2015.4} .
\bcomment{Publisher: Nature Publishing Group}.
Accessed 2024-06-03
\end{barticle}
\endbibitem

\bibitem[\protect\citeauthoryear{Chen et~al.}{2020}]{chen_critical_2020}
\begin{barticle}
\bauthor{\bsnm{Chen}, \binits{C.}},
\bauthor{\bsnm{Zuo}, \binits{Y.}},
\bauthor{\bsnm{Ye}, \binits{W.}},
\bauthor{\bsnm{Li}, \binits{X.}},
\bauthor{\bsnm{Deng}, \binits{Z.}},
\bauthor{\bsnm{Ong}, \binits{S.P.}}:
\batitle{A {Critical} {Review} of {Machine} {Learning} of {Energy} {Materials}}.
\bjtitle{Advanced Energy Materials}
\bvolume{10}(\bissue{8}),
\bfpage{1903242}
(\byear{2020})
\doiurl{10.1002/aenm.201903242} .
Accessed 2024-06-03
\end{barticle}
\endbibitem

\bibitem[\protect\citeauthoryear{T. Szczypiński et~al.}{2021}]{tszczypinski_can_2021}
\begin{barticle}
\bauthor{\bsnm{T. Szczypiński}, \binits{F.}},
\bauthor{\bsnm{Bennett}, \binits{S.}},
\bauthor{\bsnm{E. Jelfs}, \binits{K.}}:
\batitle{Can we predict materials that can be synthesised?}
\bjtitle{Chemical Science}
\bvolume{12}(\bissue{3}),
\bfpage{830}--\blpage{840}
(\byear{2021})
\doiurl{10.1039/D0SC04321D} .
\bcomment{Publisher: Royal Society of Chemistry}.
Accessed 2024-06-03
\end{barticle}
\endbibitem

\bibitem[\protect\citeauthoryear{Frenkel and Smit}{2001}]{frenkel_understanding_2001}
\begin{bbook}
\bauthor{\bsnm{Frenkel}, \binits{D.}},
\bauthor{\bsnm{Smit}, \binits{B.}}:
\bbtitle{Understanding {Molecular} {Simulation}: {From} {Algorithms} to {Applications}},
\bedition{2nd edition} edn.
\bpublisher{Academic Press},
\blocation{San Diego}
(\byear{2001})
\end{bbook}
\endbibitem

\bibitem[\protect\citeauthoryear{Hellman et~al.}{2011}]{hellman_lattice_2011}
\begin{barticle}
\bauthor{\bsnm{Hellman}, \binits{O.}},
\bauthor{\bsnm{Abrikosov}, \binits{I.A.}},
\bauthor{\bsnm{Simak}, \binits{S.I.}}:
\batitle{Lattice dynamics of anharmonic solids from first principles}.
\bjtitle{Physical Review B}
\bvolume{84}(\bissue{18}),
\bfpage{180301}
(\byear{2011})
\doiurl{10.1103/PhysRevB.84.180301} .
\bcomment{Publisher: American Physical Society}.
Accessed 2024-05-15
\end{barticle}
\endbibitem

\bibitem[\protect\citeauthoryear{Hellman et~al.}{2013}]{hellman_temperature_2013}
\begin{barticle}
\bauthor{\bsnm{Hellman}, \binits{O.}},
\bauthor{\bsnm{Steneteg}, \binits{P.}},
\bauthor{\bsnm{Abrikosov}, \binits{I.A.}},
\bauthor{\bsnm{Simak}, \binits{S.I.}}:
\batitle{Temperature dependent effective potential method for accurate free energy calculations of solids}.
\bjtitle{Physical Review B}
\bvolume{87}(\bissue{10}),
\bfpage{104111}
(\byear{2013})
\doiurl{10.1103/PhysRevB.87.104111} .
Accessed 2024-03-27
\end{barticle}
\endbibitem

\bibitem[\protect\citeauthoryear{Korotaev et~al.}{2018}]{korotaev_reproducibility_2018}
\begin{barticle}
\bauthor{\bsnm{Korotaev}, \binits{P.}},
\bauthor{\bsnm{Belov}, \binits{M.}},
\bauthor{\bsnm{Yanilkin}, \binits{A.}}:
\batitle{Reproducibility of vibrational free energy by different methods}.
\bjtitle{Computational Materials Science}
\bvolume{150},
\bfpage{47}--\blpage{53}
(\byear{2018})
\doiurl{10.1016/j.commatsci.2018.03.057} .
Accessed 2024-05-04
\end{barticle}
\endbibitem

\bibitem[\protect\citeauthoryear{Grabowski et~al.}{2019}]{grabowski_ab_2019}
\begin{barticle}
\bauthor{\bsnm{Grabowski}, \binits{B.}},
\bauthor{\bsnm{Ikeda}, \binits{Y.}},
\bauthor{\bsnm{Srinivasan}, \binits{P.}},
\bauthor{\bsnm{Körmann}, \binits{F.}},
\bauthor{\bsnm{Freysoldt}, \binits{C.}},
\bauthor{\bsnm{Duff}, \binits{A.I.}},
\bauthor{\bsnm{Shapeev}, \binits{A.}},
\bauthor{\bsnm{Neugebauer}, \binits{J.}}:
\batitle{Ab initio vibrational free energies including anharmonicity for multicomponent alloys}.
\bjtitle{npj Computational Materials}
\bvolume{5}(\bissue{1}),
\bfpage{1}--\blpage{6}
(\byear{2019})
\doiurl{10.1038/s41524-019-0218-8} .
\bcomment{Number: 1 Publisher: Nature Publishing Group}.
Accessed 2023-11-06
\end{barticle}
\endbibitem

\bibitem[\protect\citeauthoryear{Behler and Csányi}{2021}]{behler_machine_2021}
\begin{barticle}
\bauthor{\bsnm{Behler}, \binits{J.}},
\bauthor{\bsnm{Csányi}, \binits{G.}}:
\batitle{Machine learning potentials for extended systems: a perspective}.
\bjtitle{The European Physical Journal B}
\bvolume{94}(\bissue{7}),
\bfpage{142}
(\byear{2021})
\doiurl{10.1140/epjb/s10051-021-00156-1} .
Accessed 2024-05-13
\end{barticle}
\endbibitem

\bibitem[\protect\citeauthoryear{Martin}{2008}]{martin_electronic_2008}
\begin{bbook}
\bauthor{\bsnm{Martin}, \binits{R.M.}}:
\bbtitle{Electronic {Structure}: {Basic} {Theory} and {Practical} {Methods}},
\bedition{1st edition} edn.
\bpublisher{Cambridge University Press},
\blocation{Cambridge, UK ; New York}
(\byear{2008})
\end{bbook}
\endbibitem

\bibitem[\protect\citeauthoryear{Kresse and Furthmüller}{1996}]{kresse_efficiency_1996}
\begin{barticle}
\bauthor{\bsnm{Kresse}, \binits{G.}},
\bauthor{\bsnm{Furthmüller}, \binits{J.}}:
\batitle{Efficiency of ab-initio total energy calculations for metals and semiconductors using a plane-wave basis set}.
\bjtitle{Computational Materials Science}
\bvolume{6}(\bissue{1}),
\bfpage{15}--\blpage{50}
(\byear{1996})
\doiurl{10.1016/0927-0256(96)00008-0} .
Accessed 2023-04-03
\end{barticle}
\endbibitem

\bibitem[\protect\citeauthoryear{Kocer et~al.}{2022}]{kocer_neural_2022}
\begin{barticle}
\bauthor{\bsnm{Kocer}, \binits{E.}},
\bauthor{\bsnm{Ko}, \binits{T.W.}},
\bauthor{\bsnm{Behler}, \binits{J.}}:
\batitle{Neural {Network} {Potentials}: {A} {Concise} {Overview} of {Methods}}.
\bjtitle{Annual Review of Physical Chemistry}
\bvolume{73}(\bissue{Volume 73, 2022}),
\bfpage{163}--\blpage{186}
(\byear{2022})
\doiurl{10.1146/annurev-physchem-082720-034254} .
\bcomment{Publisher: Annual Reviews}.
Accessed 2024-05-13
\end{barticle}
\endbibitem

\bibitem[\protect\citeauthoryear{Liu et~al.}{2023}]{liu_discrepancies_2023}
\begin{barticle}
\bauthor{\bsnm{Liu}, \binits{Y.}},
\bauthor{\bsnm{He}, \binits{X.}},
\bauthor{\bsnm{Mo}, \binits{Y.}}:
\batitle{Discrepancies and error evaluation metrics for machine learning interatomic potentials}.
\bjtitle{npj Computational Materials}
\bvolume{9}(\bissue{1}),
\bfpage{1}--\blpage{13}
(\byear{2023})
\doiurl{10.1038/s41524-023-01123-3} .
\bcomment{Publisher: Nature Publishing Group}.
Accessed 2024-06-03
\end{barticle}
\endbibitem

\bibitem[\protect\citeauthoryear{Vita and Schwalbe-Koda}{2023}]{vita_data_2023}
\begin{barticle}
\bauthor{\bsnm{Vita}, \binits{J.A.}},
\bauthor{\bsnm{Schwalbe-Koda}, \binits{D.}}:
\batitle{Data efficiency and extrapolation trends in neural network interatomic potentials}.
\bjtitle{Machine Learning: Science and Technology}
\bvolume{4}(\bissue{3}),
\bfpage{035031}
(\byear{2023})
\doiurl{10.1088/2632-2153/acf115} .
\bcomment{Publisher: IOP Publishing}.
Accessed 2024-06-03
\end{barticle}
\endbibitem

\bibitem[\protect\citeauthoryear{Sluss et~al.}{2022}]{sluss_exploration_2022}
\begin{barticle}
\bauthor{\bsnm{Sluss}, \binits{C.C.}},
\bauthor{\bsnm{Pittman}, \binits{J.}},
\bauthor{\bsnm{Nicholson}, \binits{D.M.}},
\bauthor{\bsnm{Keffer}, \binits{D.J.}}:
\batitle{Exploration of {Entropy} {Pair} {Functional} {Theory}}.
\bjtitle{Entropy}
\bvolume{24}(\bissue{5}),
\bfpage{603}
(\byear{2022})
\doiurl{10.3390/e24050603} .
\bcomment{Number: 5 Publisher: Multidisciplinary Digital Publishing Institute}.
Accessed 2024-06-12
\end{barticle}
\endbibitem

\bibitem[\protect\citeauthoryear{Wang et~al.}{2023}]{wang_data-driven_2023}
\begin{barticle}
\bauthor{\bsnm{Wang}, \binits{X.}},
\bauthor{\bsnm{Wang}, \binits{Z.}},
\bauthor{\bsnm{Gao}, \binits{P.}},
\bauthor{\bsnm{Zhang}, \binits{C.}},
\bauthor{\bsnm{Lv}, \binits{J.}},
\bauthor{\bsnm{Wang}, \binits{H.}},
\bauthor{\bsnm{Liu}, \binits{H.}},
\bauthor{\bsnm{Wang}, \binits{Y.}},
\bauthor{\bsnm{Ma}, \binits{Y.}}:
\batitle{Data-driven prediction of complex crystal structures of dense lithium}.
\bjtitle{Nature Communications}
\bvolume{14}(\bissue{1}),
\bfpage{2924}
(\byear{2023})
\doiurl{10.1038/s41467-023-38650-y} .
\bcomment{Number: 1 Publisher: Nature Publishing Group}.
Accessed 2023-06-20
\end{barticle}
\endbibitem

\bibitem[\protect\citeauthoryear{Fultz}{2010}]{fultz_vibrational_2010}
\begin{barticle}
\bauthor{\bsnm{Fultz}, \binits{B.}}:
\batitle{Vibrational thermodynamics of materials}.
\bjtitle{Progress in Materials Science}
\bvolume{55}(\bissue{4}),
\bfpage{247}--\blpage{352}
(\byear{2010})
\doiurl{10.1016/j.pmatsci.2009.05.002} .
Accessed 2024-05-18
\end{barticle}
\endbibitem

\bibitem[\protect\citeauthoryear{Allen}{2020}]{allen_theory_2020}
\begin{barticle}
\bauthor{\bsnm{Allen}, \binits{P.B.}}:
\batitle{Theory of thermal expansion: {Quasi}-harmonic approximation and corrections from quasi-particle renormalization}.
\bjtitle{Modern Physics Letters B}
\bvolume{34}(\bissue{02}),
\bfpage{2050025}
(\byear{2020})
\doiurl{10.1142/S0217984920500256} .
\bcomment{Publisher: World Scientific Publishing Co.}
Accessed 2024-06-03
\end{barticle}
\endbibitem

\bibitem[\protect\citeauthoryear{Jerabek et~al.}{2022}]{jerabek_solving_2022}
\begin{barticle}
\bauthor{\bsnm{Jerabek}, \binits{P.}},
\bauthor{\bsnm{Burrows}, \binits{A.}},
\bauthor{\bsnm{Schwerdtfeger}, \binits{P.}}:
\batitle{Solving a problem with a single parameter: {A} smooth bcc to fcc phase transition for metallic lithium}.
\bjtitle{Chemical Communications}
\bvolume{58}(\bissue{96}),
\bfpage{13369}--\blpage{13372}
(\byear{2022})
\doiurl{10.1039/D2CC04928G} .
\bcomment{arXiv:2211.06317 [cond-mat]}.
Accessed 2024-04-19
\end{barticle}
\endbibitem

\bibitem[\protect\citeauthoryear{Huang and Widom}{2022}]{huang_vibrational_2022}
\begin{barticle}
\bauthor{\bsnm{Huang}, \binits{Y.}},
\bauthor{\bsnm{Widom}, \binits{M.}}:
\batitle{Vibrational {Entropy} of {Crystalline} {Solids} from {Covariance} of {Atomic} {Displacements}}.
\bjtitle{Entropy}
\bvolume{24}(\bissue{5}),
\bfpage{618}
(\byear{2022})
\doiurl{10.3390/e24050618} .
\bcomment{Number: 5 Publisher: Multidisciplinary Digital Publishing Institute}.
Accessed 2023-04-25
\end{barticle}
\endbibitem

\bibitem[\protect\citeauthoryear{Kadkhodaei et~al.}{2017}]{kadkhodaei_free_2017}
\begin{barticle}
\bauthor{\bsnm{Kadkhodaei}, \binits{S.}},
\bauthor{\bsnm{Hong}, \binits{Q.-J.}},
\bauthor{\bsnm{Van De~Walle}, \binits{A.}}:
\batitle{Free energy calculation of mechanically unstable but dynamically stabilized bcc titanium}.
\bjtitle{Physical Review B}
\bvolume{95}(\bissue{6}),
\bfpage{064101}
(\byear{2017})
\doiurl{10.1103/PhysRevB.95.064101} .
Accessed 2024-06-11
\end{barticle}
\endbibitem

\bibitem[\protect\citeauthoryear{Romero et~al.}{2015}]{romero_thermal_2015}
\begin{barticle}
\bauthor{\bsnm{Romero}, \binits{A.H.}},
\bauthor{\bsnm{Gross}, \binits{E.K.U.}},
\bauthor{\bsnm{Verstraete}, \binits{M.J.}},
\bauthor{\bsnm{Hellman}, \binits{O.}}:
\batitle{Thermal conductivity in {PbTe} from first principles}.
\bjtitle{Physical Review B}
\bvolume{91}(\bissue{21}),
\bfpage{214310}
(\byear{2015})
\doiurl{10.1103/PhysRevB.91.214310} .
Accessed 2024-05-22
\end{barticle}
\endbibitem

\bibitem[\protect\citeauthoryear{Phuthi et~al.}{2024}]{phuthi_accurate_2024}
\begin{barticle}
\bauthor{\bsnm{Phuthi}, \binits{M.K.}},
\bauthor{\bsnm{Yao}, \binits{A.M.}},
\bauthor{\bsnm{Batzner}, \binits{S.}},
\bauthor{\bsnm{Musaelian}, \binits{A.}},
\bauthor{\bsnm{Guan}, \binits{P.}},
\bauthor{\bsnm{Kozinsky}, \binits{B.}},
\bauthor{\bsnm{Cubuk}, \binits{E.D.}},
\bauthor{\bsnm{Viswanathan}, \binits{V.}}:
\batitle{Accurate {Surface} and {Finite}-{Temperature} {Bulk} {Properties} of {Lithium} {Metal} at {Large} {Scales} {Using} {Machine} {Learning} {Interaction} {Potentials}}.
\bjtitle{ACS Omega}
\bvolume{9}(\bissue{9}),
\bfpage{10904}--\blpage{10912}
(\byear{2024})
\doiurl{10.1021/acsomega.3c10014} .
\bcomment{Publisher: American Chemical Society}.
Accessed 2024-03-15
\end{barticle}
\endbibitem

\bibitem[\protect\citeauthoryear{Vandermause et~al.}{2020}]{vandermause_--fly_2020}
\begin{barticle}
\bauthor{\bsnm{Vandermause}, \binits{J.}},
\bauthor{\bsnm{Torrisi}, \binits{S.B.}},
\bauthor{\bsnm{Batzner}, \binits{S.}},
\bauthor{\bsnm{Xie}, \binits{Y.}},
\bauthor{\bsnm{Sun}, \binits{L.}},
\bauthor{\bsnm{Kolpak}, \binits{A.M.}},
\bauthor{\bsnm{Kozinsky}, \binits{B.}}:
\batitle{On-the-fly active learning of interpretable {Bayesian} force fields for atomistic rare events}.
\bjtitle{npj Computational Materials}
\bvolume{6}(\bissue{1}),
\bfpage{1}--\blpage{11}
(\byear{2020})
\doiurl{10.1038/s41524-020-0283-z} .
\bcomment{Publisher: Nature Publishing Group}.
Accessed 2024-06-08
\end{barticle}
\endbibitem

\bibitem[\protect\citeauthoryear{Deng et~al.}{2023}]{deng_chgnet_2023}
\begin{barticle}
\bauthor{\bsnm{Deng}, \binits{B.}},
\bauthor{\bsnm{Zhong}, \binits{P.}},
\bauthor{\bsnm{Jun}, \binits{K.}},
\bauthor{\bsnm{Riebesell}, \binits{J.}},
\bauthor{\bsnm{Han}, \binits{K.}},
\bauthor{\bsnm{Bartel}, \binits{C.J.}},
\bauthor{\bsnm{Ceder}, \binits{G.}}:
\batitle{{CHGNet} as a pretrained universal neural network potential for charge-informed atomistic modelling}.
\bjtitle{Nature Machine Intelligence}
\bvolume{5}(\bissue{9}),
\bfpage{1031}--\blpage{1041}
(\byear{2023})
\doiurl{10.1038/s42256-023-00716-3} .
\bcomment{Publisher: Nature Publishing Group}.
Accessed 2024-06-08
\end{barticle}
\endbibitem

\bibitem[\protect\citeauthoryear{Batatia et~al.}{2022}]{batatia_mace_2022}
\begin{bchapter}
\bauthor{\bsnm{Batatia}, \binits{I.}},
\bauthor{\bsnm{Kovacs}, \binits{D.P.}},
\bauthor{\bsnm{Simm}, \binits{G.N.C.}},
\bauthor{\bsnm{Ortner}, \binits{C.}},
\bauthor{\bsnm{Csanyi}, \binits{G.}}:
\bctitle{{MACE}: {Higher} {Order} {Equivariant} {Message} {Passing} {Neural} {Networks} for {Fast} and {Accurate} {Force} {Fields}}.
(\byear{2022}).
\burl{https://openreview.net/forum?id=YPpSngE-ZU}
Accessed 2024-06-08
\end{bchapter}
\endbibitem

\bibitem[\protect\citeauthoryear{Batzner et~al.}{2022}]{batzner_e3-equivariant_2022}
\begin{barticle}
\bauthor{\bsnm{Batzner}, \binits{S.}},
\bauthor{\bsnm{Musaelian}, \binits{A.}},
\bauthor{\bsnm{Sun}, \binits{L.}},
\bauthor{\bsnm{Geiger}, \binits{M.}},
\bauthor{\bsnm{Mailoa}, \binits{J.P.}},
\bauthor{\bsnm{Kornbluth}, \binits{M.}},
\bauthor{\bsnm{Molinari}, \binits{N.}},
\bauthor{\bsnm{Smidt}, \binits{T.E.}},
\bauthor{\bsnm{Kozinsky}, \binits{B.}}:
\batitle{E(3)-equivariant graph neural networks for data-efficient and accurate interatomic potentials}.
\bjtitle{Nature Communications}
\bvolume{13}(\bissue{1}),
\bfpage{2453}
(\byear{2022})
\doiurl{10.1038/s41467-022-29939-5} .
\bcomment{Number: 1 Publisher: Nature Publishing Group}.
Accessed 2022-06-20
\end{barticle}
\endbibitem

\bibitem[\protect\citeauthoryear{Chuan-Hui and Xiao-Bao}{2001}]{chuan-hui_gruneisen_2001}
\begin{barticle}
\bauthor{\bsnm{Chuan-Hui}, \binits{N.}},
\bauthor{\bsnm{Xiao-Bao}, \binits{S.}}:
\batitle{Grüneisen parameter of lithium, sodium and potassium at \textit{{T}}=298 {K}}.
\bjtitle{Journal of Physics and Chemistry of Solids}
\bvolume{62}(\bissue{7}),
\bfpage{1359}--\blpage{1361}
(\byear{2001})
\doiurl{10.1016/S0022-3697(01)00007-5} .
Accessed 2024-05-18
\end{barticle}
\endbibitem

\bibitem[\protect\citeauthoryear{Beg and Nielsen}{1976}]{beg_temperature_1976}
\begin{barticle}
\bauthor{\bsnm{Beg}, \binits{M.M.}},
\bauthor{\bsnm{Nielsen}, \binits{M.}}:
\batitle{Temperature dependence of lattice dynamics of lithium 7}.
\bjtitle{Physical Review B}
\bvolume{14}(\bissue{10}),
\bfpage{4266}--\blpage{4273}
(\byear{1976})
\doiurl{10.1103/PhysRevB.14.4266} .
\bcomment{Publisher: American Physical Society}.
Accessed 2022-05-10
\end{barticle}
\endbibitem

\bibitem[\protect\citeauthoryear{Owen and Williams}{1954}]{owen_x-ray_1954}
\begin{barticle}
\bauthor{\bsnm{Owen}, \binits{E.A.}},
\bauthor{\bsnm{Williams}, \binits{G.I.}}:
\batitle{X-{Ray} {Measurements} on {Lithium} at {Low} {Temperatures}}.
\bjtitle{Proceedings of the Physical Society. Section A}
\bvolume{67}(\bissue{10}),
\bfpage{895}--\blpage{900}
(\byear{1954})
\doiurl{10.1088/0370-1298/67/10/306} .
Accessed 2022-11-10
\end{barticle}
\endbibitem

\bibitem[\protect\citeauthoryear{Chase}{1998}]{chase_nist-janaf_1998}
\begin{botherref}
\oauthor{\bsnm{Chase}, \binits{M.}}:
Nist-janaf thermochemical tables, 4th edition
(1998)
\end{botherref}
\endbibitem

\bibitem[\protect\citeauthoryear{Douglas et~al.}{1955}]{douglas_lithium_1955}
\begin{barticle}
\bauthor{\bsnm{Douglas}, \binits{T.B.}},
\bauthor{\bsnm{Epstein}, \binits{L.F.}},
\bauthor{\bsnm{Dever}, \binits{J.L.}},
\bauthor{\bsnm{Howland}, \binits{W.H.}}:
\batitle{Lithium: {Heat} {Content} from 0 to 900°, {Triple} {Point} and {Heat} of {Fusion}, and {Thermodynamic} {Properties} of the {Solid} and {Liquid1}}.
\bjtitle{Journal of the American Chemical Society}
\bvolume{77}(\bissue{8}),
\bfpage{2144}--\blpage{2150}
(\byear{1955})
\doiurl{10.1021/ja01613a031} .
\bcomment{Publisher: American Chemical Society}.
Accessed 2024-06-07
\end{barticle}
\endbibitem

\bibitem[\protect\citeauthoryear{Sholl}{2009}]{sholl_density_2009}
\begin{bbook}
\bauthor{\bsnm{Sholl}, \binits{D.S.}}:
\bbtitle{Density Functional Theory a Practical Introduction}.
\bpublisher{Wiley},
\blocation{Hoboken, N.J}
(\byear{2009})
\end{bbook}
\endbibitem

\bibitem[\protect\citeauthoryear{Markland and Ceriotti}{2018}]{markland_nuclear_2018}
\begin{barticle}
\bauthor{\bsnm{Markland}, \binits{T.E.}},
\bauthor{\bsnm{Ceriotti}, \binits{M.}}:
\batitle{Nuclear quantum effects enter the mainstream}.
\bjtitle{Nature Reviews Chemistry}
\bvolume{2}(\bissue{3}),
\bfpage{1}--\blpage{14}
(\byear{2018})
\doiurl{10.1038/s41570-017-0109} .
\bcomment{Publisher: Nature Publishing Group}.
Accessed 2024-05-21
\end{barticle}
\endbibitem

\bibitem[\protect\citeauthoryear{Wigner}{1932}]{wigner_quantum_1932}
\begin{barticle}
\bauthor{\bsnm{Wigner}, \binits{E.}}:
\batitle{On the {Quantum} {Correction} {For} {Thermodynamic} {Equilibrium}}.
\bjtitle{Physical Review}
\bvolume{40}(\bissue{5}),
\bfpage{749}--\blpage{759}
(\byear{1932})
\doiurl{10.1103/PhysRev.40.749} .
\bcomment{Publisher: American Physical Society}.
Accessed 2024-06-03
\end{barticle}
\endbibitem

\bibitem[\protect\citeauthoryear{Deserno and Turgut}{2023}]{deserno_leading_2023}
\begin{barticle}
\bauthor{\bsnm{Deserno}, \binits{M.}},
\bauthor{\bsnm{Turgut}, \binits{O.T.}}:
\batitle{Leading quantum correction to the classical free energy}.
\bjtitle{American Journal of Physics}
\bvolume{91}(\bissue{11}),
\bfpage{923}--\blpage{931}
(\byear{2023})
\doiurl{10.1119/5.0106687} .
Accessed 2024-06-03
\end{barticle}
\endbibitem

\bibitem[\protect\citeauthoryear{Huang et~al.}{2022}]{huang_ab_2022}
\begin{barticle}
\bauthor{\bsnm{Huang}, \binits{Y.}},
\bauthor{\bsnm{Widom}, \binits{M.}},
\bauthor{\bsnm{Gao}, \binits{M.C.}}:
\batitle{\textit{{Ab} initio} free energies of liquid metal alloys: {Application} to the phase diagrams of {Li}-{Na} and {K}-{Na}}.
\bjtitle{Physical Review Materials}
\bvolume{6}(\bissue{1}),
\bfpage{013802}
(\byear{2022})
\doiurl{10.1103/PhysRevMaterials.6.013802} .
Accessed 2023-11-16
\end{barticle}
\endbibitem

\bibitem[\protect\citeauthoryear{Hultgren}{1963}]{hultgren_selected_1963}
\begin{bbook}
\bauthor{\bsnm{Hultgren}, \binits{R.}}:
\bbtitle{Selected {Values} of {Thermodynamic} {Properties} of {Metals} And {Alloys}}.
\bpublisher{Wiley}, \blocation{???}
(\byear{1963}).
\bcomment{Google-Books-ID: AjFRAAAAMAAJ}
\end{bbook}
\endbibitem

\bibitem[\protect\citeauthoryear{Xie et~al.}{2008}]{xie_origin_2008}
\begin{barticle}
\bauthor{\bsnm{Xie}, \binits{Y.}},
\bauthor{\bsnm{Ma}, \binits{Y.M.}},
\bauthor{\bsnm{Cui}, \binits{T.}},
\bauthor{\bsnm{Li}, \binits{Y.}},
\bauthor{\bsnm{Qiu}, \binits{J.}},
\bauthor{\bsnm{Zou}, \binits{G.T.}}:
\batitle{Origin of bcc to fcc phase transition under pressure in alkali metals}.
\bjtitle{New Journal of Physics}
\bvolume{10}(\bissue{6}),
\bfpage{063022}
(\byear{2008})
\doiurl{10.1088/1367-2630/10/6/063022} .
Accessed 2024-04-19
\end{barticle}
\endbibitem

\bibitem[\protect\citeauthoryear{Schaeffer et~al.}{2015}]{schaeffer_boundaries_2015}
\begin{barticle}
\bauthor{\bsnm{Schaeffer}, \binits{A.M.}},
\bauthor{\bsnm{Cai}, \binits{W.}},
\bauthor{\bsnm{Olejnik}, \binits{E.}},
\bauthor{\bsnm{Molaison}, \binits{J.J.}},
\bauthor{\bsnm{Sinogeikin}, \binits{S.}},
\bauthor{\bsnm{Santos}, \binits{A.M.}},
\bauthor{\bsnm{Deemyad}, \binits{S.}}:
\batitle{Boundaries for martensitic transition of {7Li} under pressure}.
\bjtitle{Nature Communications}
\bvolume{6}(\bissue{1}),
\bfpage{8030}
(\byear{2015})
\doiurl{10.1038/ncomms9030} .
\bcomment{Number: 1 Publisher: Nature Publishing Group}.
Accessed 2023-09-27
\end{barticle}
\endbibitem

\bibitem[\protect\citeauthoryear{Liu et~al.}{2022}]{liu_zentropy_2022}
\begin{barticle}
\bauthor{\bsnm{Liu}, \binits{Z.-K.}},
\bauthor{\bsnm{Wang}, \binits{Y.}},
\bauthor{\bsnm{Shang}, \binits{S.-L.}}:
\batitle{Zentropy {Theory} for {Positive} and {Negative} {Thermal} {Expansion}}.
\bjtitle{Journal of Phase Equilibria and Diffusion}
\bvolume{43}(\bissue{6}),
\bfpage{598}--\blpage{605}
(\byear{2022})
\doiurl{10.1007/s11669-022-00942-z} .
Accessed 2024-06-12
\end{barticle}
\endbibitem

\bibitem[\protect\citeauthoryear{Frenkel and Smit}{2002}]{frenkel_appendix_2002}
\begin{botherref}
Appendix {H} - {Statistical} {Mechanics} of the {Gibbs} “{Ensemble}”.
In: \oauthor{\bsnm{Frenkel}, \binits{D.}},
\oauthor{\bsnm{Smit}, \binits{B.}} (eds.)
Understanding {Molecular} {Simulation} ({Second} {Edition}),
pp. 563--571.
Academic Press,
San Diego
(2002).
\doiurl{10.1016/B978-012267351-1/50027-4} .
\url{https://www.sciencedirect.com/science/article/pii/B9780122673511500274}
Accessed 2024-05-18
\end{botherref}
\endbibitem

\bibitem[\protect\citeauthoryear{Giannozzi et~al.}{2009}]{giannozzi_quantum_2009}
\begin{barticle}
\bauthor{\bsnm{Giannozzi}, \binits{P.}},
\bauthor{\bsnm{Baroni}, \binits{S.}},
\bauthor{\bsnm{Bonini}, \binits{N.}},
\bauthor{\bsnm{Calandra}, \binits{M.}},
\bauthor{\bsnm{Car}, \binits{R.}},
\bauthor{\bsnm{Cavazzoni}, \binits{C.}},
\bauthor{\bsnm{Ceresoli}, \binits{D.}},
\bauthor{\bsnm{Chiarotti}, \binits{G.L.}},
\bauthor{\bsnm{Cococcioni}, \binits{M.}},
\bauthor{\bsnm{Dabo}, \binits{I.}},
\bauthor{\bsnm{Corso}, \binits{A.D.}},
\bauthor{\bsnm{Gironcoli}, \binits{S.d.}},
\bauthor{\bsnm{Fabris}, \binits{S.}},
\bauthor{\bsnm{Fratesi}, \binits{G.}},
\bauthor{\bsnm{Gebauer}, \binits{R.}},
\bauthor{\bsnm{Gerstmann}, \binits{U.}},
\bauthor{\bsnm{Gougoussis}, \binits{C.}},
\bauthor{\bsnm{Kokalj}, \binits{A.}},
\bauthor{\bsnm{Lazzeri}, \binits{M.}},
\bauthor{\bsnm{Martin-Samos}, \binits{L.}},
\bauthor{\bsnm{Marzari}, \binits{N.}},
\bauthor{\bsnm{Mauri}, \binits{F.}},
\bauthor{\bsnm{Mazzarello}, \binits{R.}},
\bauthor{\bsnm{Paolini}, \binits{S.}},
\bauthor{\bsnm{Pasquarello}, \binits{A.}},
\bauthor{\bsnm{Paulatto}, \binits{L.}},
\bauthor{\bsnm{Sbraccia}, \binits{C.}},
\bauthor{\bsnm{Scandolo}, \binits{S.}},
\bauthor{\bsnm{Sclauzero}, \binits{G.}},
\bauthor{\bsnm{Seitsonen}, \binits{A.P.}},
\bauthor{\bsnm{Smogunov}, \binits{A.}},
\bauthor{\bsnm{Umari}, \binits{P.}},
\bauthor{\bsnm{Wentzcovitch}, \binits{R.M.}}:
\batitle{{QUANTUM} {ESPRESSO}: a modular and open-source software project for quantum simulations of materials}.
\bjtitle{Journal of Physics: Condensed Matter}
\bvolume{21}(\bissue{39}),
\bfpage{395502}
(\byear{2009})
\doiurl{10.1088/0953-8984/21/39/395502} .
Accessed 2023-02-27
\end{barticle}
\endbibitem

\bibitem[\protect\citeauthoryear{Perdew et~al.}{1996}]{perdew_generalized_1996}
\begin{barticle}
\bauthor{\bsnm{Perdew}, \binits{J.P.}},
\bauthor{\bsnm{Burke}, \binits{K.}},
\bauthor{\bsnm{Ernzerhof}, \binits{M.}}:
\batitle{Generalized {Gradient} {Approximation} {Made} {Simple}}.
\bjtitle{Physical Review Letters}
\bvolume{77}(\bissue{18}),
\bfpage{3865}--\blpage{3868}
(\byear{1996})
\doiurl{10.1103/PhysRevLett.77.3865} .
Accessed 2019-12-03
\end{barticle}
\endbibitem

\bibitem[\protect\citeauthoryear{Blöchl}{1994}]{blochl_projector_1994}
\begin{barticle}
\bauthor{\bsnm{Blöchl}, \binits{P.E.}}:
\batitle{Projector augmented-wave method}.
\bjtitle{Physical Review B}
\bvolume{50}(\bissue{24}),
\bfpage{17953}--\blpage{17979}
(\byear{1994})
\doiurl{10.1103/PhysRevB.50.17953} .
\bcomment{Publisher: American Physical Society}.
Accessed 2023-02-27
\end{barticle}
\endbibitem

\bibitem[\protect\citeauthoryear{Monkhorst and Pack}{1976}]{monkhorst_special_1976}
\begin{barticle}
\bauthor{\bsnm{Monkhorst}, \binits{H.J.}},
\bauthor{\bsnm{Pack}, \binits{J.D.}}:
\batitle{Special points for {Brillouin}-zone integrations}.
\bjtitle{Physical Review B}
\bvolume{13}(\bissue{12}),
\bfpage{5188}--\blpage{5192}
(\byear{1976})
\doiurl{10.1103/PhysRevB.13.5188} .
Accessed 2021-02-16
\end{barticle}
\endbibitem

\bibitem[\protect\citeauthoryear{Methfessel and Paxton}{1989}]{methfessel_high-precision_1989}
\begin{barticle}
\bauthor{\bsnm{Methfessel}, \binits{M.}},
\bauthor{\bsnm{Paxton}, \binits{A.T.}}:
\batitle{High-precision sampling for {Brillouin}-zone integration in metals}.
\bjtitle{Physical Review B}
\bvolume{40}(\bissue{6}),
\bfpage{3616}--\blpage{3621}
(\byear{1989})
\doiurl{10.1103/PhysRevB.40.3616} .
\bcomment{Publisher: American Physical Society}.
Accessed 2023-02-27
\end{barticle}
\endbibitem

\bibitem[\protect\citeauthoryear{Thompson et~al.}{2022}]{thompson_lammps_2022}
\begin{barticle}
\bauthor{\bsnm{Thompson}, \binits{A.P.}},
\bauthor{\bsnm{Aktulga}, \binits{H.M.}},
\bauthor{\bsnm{Berger}, \binits{R.}},
\bauthor{\bsnm{Bolintineanu}, \binits{D.S.}},
\bauthor{\bsnm{Brown}, \binits{W.M.}},
\bauthor{\bsnm{Crozier}, \binits{P.S.}},
\bauthor{\bsnm{Veld}, \binits{P.J.i.t.}},
\bauthor{\bsnm{Kohlmeyer}, \binits{A.}},
\bauthor{\bsnm{Moore}, \binits{S.G.}},
\bauthor{\bsnm{Nguyen}, \binits{T.D.}},
\bauthor{\bsnm{Shan}, \binits{R.}},
\bauthor{\bsnm{Stevens}, \binits{M.J.}},
\bauthor{\bsnm{Tranchida}, \binits{J.}},
\bauthor{\bsnm{Trott}, \binits{C.}},
\bauthor{\bsnm{Plimpton}, \binits{S.J.}}:
\batitle{{LAMMPS} - a flexible simulation tool for particle-based materials modeling at the atomic, meso, and continuum scales}.
\bjtitle{Comp. Phys. Comm.}
\bvolume{271},
\bfpage{108171}
(\byear{2022})
\doiurl{10.1016/j.cpc.2021.108171}
\end{barticle}
\endbibitem

\bibitem[\protect\citeauthoryear{Shinoda et~al.}{2004}]{shinoda_rapid_2004}
\begin{barticle}
\bauthor{\bsnm{Shinoda}, \binits{W.}},
\bauthor{\bsnm{Shiga}, \binits{M.}},
\bauthor{\bsnm{Mikami}, \binits{M.}}:
\batitle{Rapid estimation of elastic constants by molecular dynamics simulation under constant stress}.
\bjtitle{Physical Review B}
\bvolume{69}(\bissue{13}),
\bfpage{134103}
(\byear{2004})
\doiurl{10.1103/PhysRevB.69.134103} .
\bcomment{Publisher: American Physical Society}.
Accessed 2024-01-10
\end{barticle}
\endbibitem

\bibitem[\protect\citeauthoryear{Mgcini}{2024}]{mgcini_dataset_2024}
\begin{botherref}
\oauthor{\bsnm{Mgcini}, \binits{P.}}:
Dataset and {Potentials} for {Lithium} {Metal}, {Phuthi} {M}. {K}. et al. (2024).
Zenodo
(2024).
\doiurl{10.5281/zenodo.10470793} .
\url{https://zenodo.org/records/10470793}
Accessed 2024-05-18
\end{botherref}
\endbibitem

\bibitem[\protect\citeauthoryear{Knoop et~al.}{2024}]{knoop_tdep_2024}
\begin{barticle}
\bauthor{\bsnm{Knoop}, \binits{F.}},
\bauthor{\bsnm{Shulumba}, \binits{N.}},
\bauthor{\bsnm{Castellano}, \binits{A.}},
\bauthor{\bsnm{Batista}, \binits{J.P.A.}},
\bauthor{\bsnm{Farris}, \binits{R.}},
\bauthor{\bsnm{Verstraete}, \binits{M.J.}},
\bauthor{\bsnm{Heine}, \binits{M.}},
\bauthor{\bsnm{Broido}, \binits{D.}},
\bauthor{\bsnm{Kim}, \binits{D.S.}},
\bauthor{\bsnm{Klarbring}, \binits{J.}},
\bauthor{\bsnm{Abrikosov}, \binits{I.A.}},
\bauthor{\bsnm{Simak}, \binits{S.I.}},
\bauthor{\bsnm{Hellman}, \binits{O.}}:
\batitle{{TDEP}: {Temperature} {Dependent} {Effective} {Potentials}}.
\bjtitle{Journal of Open Source Software}
\bvolume{9}(\bissue{94}),
\bfpage{6150}
(\byear{2024})
\doiurl{10.21105/joss.06150} .
Accessed 2024-06-01
\end{barticle}
\endbibitem

\end{thebibliography}

\pagebreak

\section*{Supplementary Information for: Vibrational Entropy and Free Energy of Solid Lithium using Covariance of Atomic Displacements Enabled by Machine Learning}

\subsection*{Consistency of CAD thermodynamics with Molecular Dynamics}

\begin{figure}[!tbh]
\centering
\includegraphics[width=\linewidth]{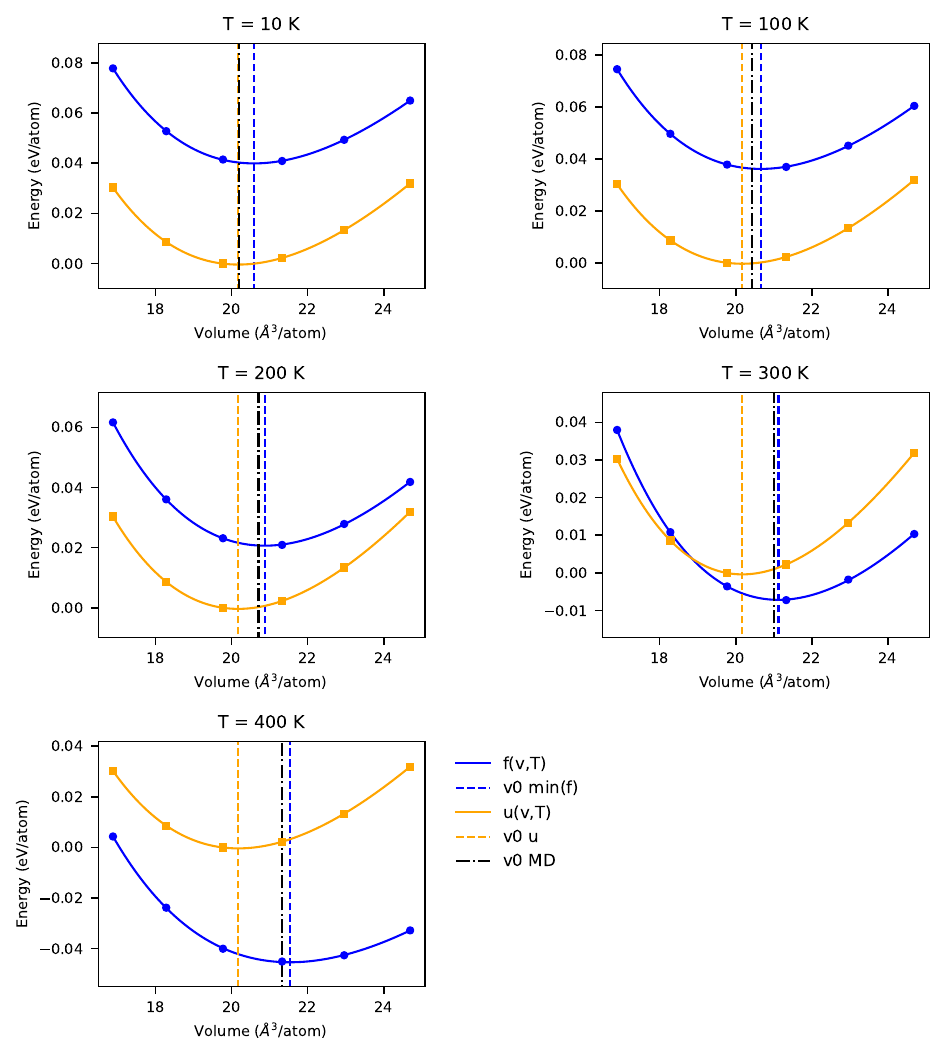}
\caption{Energy-volume curves for the static internal energy ($u(v,T=0)$) and CAD predicted Helmholtz free energy ($f(v,T)$). The CAD predicted equilibrium volume is slightly higher but consistent with the equilibrium volume found using NPT simulations in MD ($v0$)}\label{fig:minimization}
\end{figure}

In \fref{fig:minimization}, we show the energy-volume curves for the static potential energy ($u$) which is typically used as the lattice constant prediction at 0K using DFT. The lattice constant is usually determined by finding the minimum of the curve and is shown as an orange vertical line. In addition we show the Helmholtz free energy predicted by CAD as a function of volume and the corresponding minimum which includes the effect of thermal expansion and the zero-point volume correction. The lattice constant predicted using the minimum of the Helmholtz free energy is therefore always higher than that of the internal energy. One can also predict the lattice constant using NPT simulations in MD and that gives a result in between the other cases as it accounts for expansion due to kinetic energy but not the zero-point volume correction.

\begin{figure}[!tbh]
\centering
\includegraphics[width=0.5\linewidth]{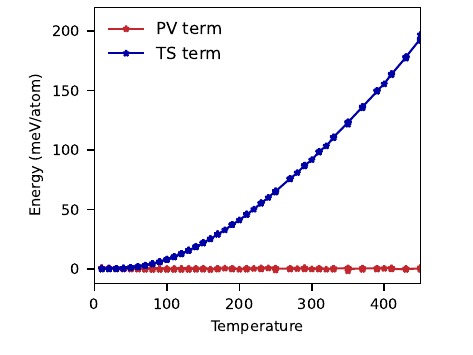}
\caption{Relative contribution of $pV$ and $Ts$ term at zero pressure as a function of temperature.}\label{fig:pvvsts}
\end{figure}

We can also do a comparison of the individual contribution of the $pV$ and $Ts$ terms in the free energies as shown in \fref{fig:pvvsts}. Since the equilibrium volumes from NPT simulations were used for the NVT simulations in CAD, the pressure and thus $pV$ are very close to zero, much smaller than the contribution of $Ts$ term from the entropy.

\begin{figure}[!tbh]
\centering
\includegraphics[width=0.5\linewidth]{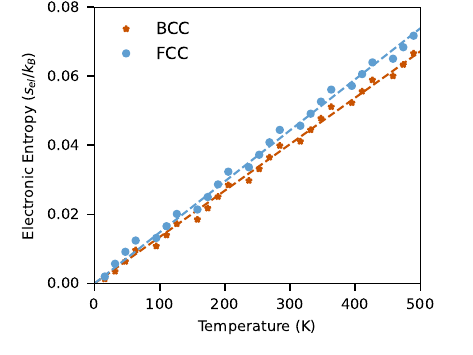}
\caption{Electronic entropy of BCC and FCC lithium as a function of temperature.}\label{fig:s_el}
\end{figure}

The electronic entropy adds a small linear addition to the vibrational entropy and is shown in \fref{fig:s_el}. The entropy of each phase is calculated by applying Fermi-Dirac smearing to static DFT calculations at the corresponding equilibrium volume. There is very little difference between the entropy values hence one would expext that this does not affect phase transitions as much. More thorough consideration might be necessary for cases where interesting electronic ground states emerge such as the formation of electrides in lithium at extreme gigapascal pressures.

\subsection*{More details on sampling strategies}
We tested a number of different sampling strategies and quantify their effects for a few examples by predicting the entropy of BCC lithium at 300 K. 

\begin{figure}[!tbh]
\centering
\includegraphics[width=0.5\linewidth]{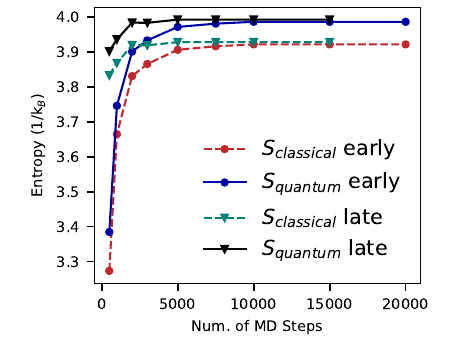}
\caption{Starting sampling later in the simulation leads to faster convergence. If storage is a limiting factor, then one can only start storing the coordinates of the trajectory after some burn-in time.}\label{fig:early_vs_late}
\end{figure}

In \fref{fig:early_vs_late}, we compare the convergence of sampling from the initial ideal crystal in the MD simulation with the case where we start sampling after the 5,000th timestep. This could be of significance if the initial configuration or velocities fluctuate at the beginning of the simulation or if computer storage for the trajectories is limited. We see that there is no significant difference in the observed results after sufficient sampling as long as the BCC structure remains stable throughout the simulation.

\begin{figure}[!tbh]
\centering
\includegraphics[width=0.5\linewidth]{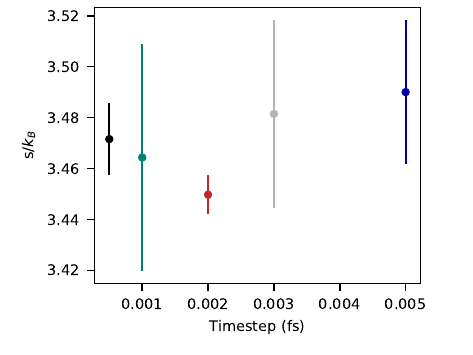}
\caption{Interestingly, the choice of timestep has very little effect on the predicted entropy with all predictions being within the small standard deviation of results. Even for relatively large timesteps of 0.005fs. For faster convergence one can therefore use a larger timestep.}\label{fig:timestep}
\end{figure}

Another important consideration in MD simulations is the timestep chosen in the simulation. The timestep is typically chosen to match with the fastest mode expected in the simulation \cite{frenkel_appendix_2002} on the order of 1fs. In \fref{fig:timestep} we show that the timestep has little effect for a small range of investigated timesteps with differences on the order of 0.02$k_B$. It therefore seems advantageous to use larger timesteps to sample more effectively and we found that this can sometimes lead to faster convergence of sensitive properties like the phonon dispersion. The effect of timestep at different temperatures would have to be considered further.

A common strategy for improving sampling is to only take samples for post-processing after every N steps where N is the stride. This makes the samples less correlated therefore leading to better statistical averages. We compare the effect of different N with the 1,000 of snapshots used in the CAD calculation. We see that the stride also does not have a significant effect by noting that the poor convergence at lower strides is actually due to not sampling for long enough. The convergence of overall length of the simulation is more significant which occurs at about 40ps.

\begin{figure}[!tbh]
\centering
\includegraphics[width=0.5\linewidth]{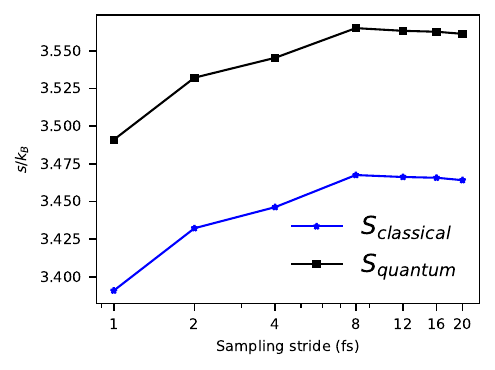}\label{fig:stride}
\caption{Convergence of the entropy with different choice of sampling stride using 1,000 timesteps for each point.}
\end{figure}

\subsection*{Inconsistency of Classical Free Energy with the Second Law of Thermodynamics at Low Temperature}
\begin{figure}[!tbh]
\centering
\includegraphics[width=0.75\linewidth]{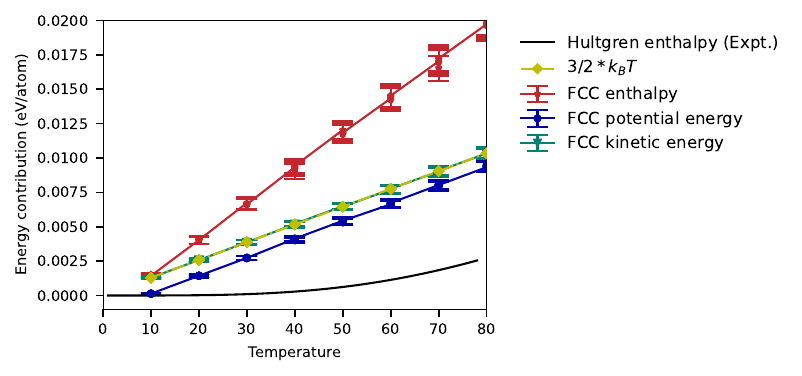}
\caption{Plot of the energy contributions to the FCC enthalpy at low temperatures. The experimental enthalpy derived from integrating the heat capacity is significantly lower than the MD predicted contributions, largely due to quantum effects.}\label{fig:md_enthalpy}
\end{figure}

As stated in the text, one might naïvely expect that calculating the Gibbs free energy using 
\begin{equation}\label{eq:uts}
    g = u - Ts + pV
\end{equation}
is valid at all temperatures where $u$ is the internal energy from MD averaging. This however leads to a gross violation of the second law of thermodynamics. This is attributed to the failure of classical Born-Oppenheimer dynamics at lower temperatures, more so for light elements where there is expected to be significant coherence and nuclear quantum effects. It is there only valid to use the quantum mechanical free energy contribution in the low temperature regime. The cause is due to the overstimation of the enthalpy as shown in \fref{fig:md_enthalpy}. The Boltzmann velocity distribution from a classical simulation does not correspond to the more coherent quantum mechanical dynamics which is dominated by the long wavelength phonon modes. In other words, the equipartition theorem does not hold.

\begin{figure}[!tbh]
\centering
\includegraphics[width=0.75\linewidth]{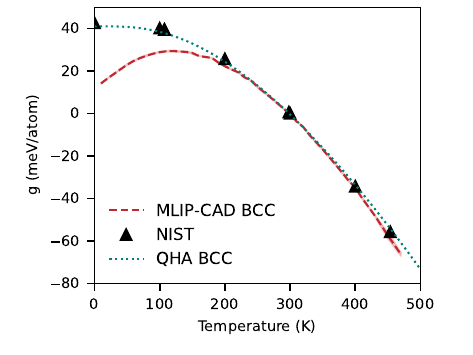}
\caption{Using eq. \ref{eq:uts} leads to a violation of the second law of thermodynamics as g is supposed to be a strictly decreasing function of T.}\label{fig:badg}
\end{figure}

\end{document}